# Seismic Inversion using Hybrid Quantum Neural Networks

**Divakar Vashisth[*1], Rohan Sharma[2,3], Tejas Ganesh Iyer[4], Tapan Mukerji[1,2,6] and Mrinal K. Sen[5]**
*[1]Department of Energy Science and Engineering, Stanford University, USA. [2]Department of Geophysics, Stanford University, USA. [3]Department of Applied Geophysics, Indian Institute of Technology (Indian School of Mines) Dhanbad, India. [4]Department of Physics, Indian Institute of Technology (Indian School of Mines) Dhanbad, India. [5]Department of Earth and Planetary Sciences and Institute for Geophysics, The University of Texas at Austin, USA. [6]Department of Earth and Planetary Sciences, Stanford University, USA*

[*]Divakar Vashisth (divakar.vashisth98@gmail.com)

## Summary

Seismic inversion-including post-stack, pre-stack, and full waveform inversion is compute and memory-intensive. Recently, several approaches, including physics-informed machine learning, have been developed to address some of these limitations. Motivated by the potential of quantum computing, we report on our attempt to map one such classical physics-informed algorithm to a quantum framework. The primary goal is to investigate the technical challenges of this mapping, given that quantum algorithms rely on computing principles fundamentally different from those in classical computing. Quantum computers operate using qubits, which exploit superposition and entanglement, offering the potential to solve classically intractable problems. While current quantum hardware is limited, hybrid quantum-classical algorithms-particularly in quantum machine learning (QML)-demonstrate potential for near-term applications and can be readily simulated. We apply QML to subsurface imaging through the development of a hybrid quantum physics-informed neural network (HQ-PINN) for post-stack and pre-stack seismic inversion. The HQ-PINN architecture adopts an encoder-decoder structure: a hybrid quantum neural network encoder estimates P- and S-impedances from seismic data, while the decoder reconstructs seismic responses using geophysical relationships. Training is guided by minimizing the misfit between the input and reconstructed seismic traces. We systematically assess the impact of quantum layer design, differentiation strategies, and simulator backends on inversion performance. We demonstrate the efficacy of our approach through the inversion of both synthetic and the Sleipner field datasets. We further evaluate the effect of noisy quantum simulators- designed to mimic real quantum hardware- on the performance and stability of the HQ-PINN model. The HQ-PINN framework consistently yields accurate results, showcasing quantum computing's promise for geosciences and paving the way for future quantum-enhanced geophysical workflows.

## Introduction

Seismic inversion plays an important role in reservoir characterization, enabling the estimation of subsurface elastic properties from surface-recorded seismic data (Tarantola, 2005; Sen, 2006; Sen & Stoffa, 2013). Traditionally, this inverse problem has been addressed using deterministic and stochastic approaches, each with distinct strengths and limitations. Deterministic methods iteratively minimize a misfit function but are often sensitive to the choice of initial models (Tarantola, 2005) and can yield non-unique solutions, necessitating the use of regularization techniques (Zhang & Castagna, 2011; Aster et al., 2018). Stochastic techniques (e.g., Zhang et al., 2012; Sen & Biswas, 2017), in contrast, use probabilistic sampling to explore the solution space, providing uncertainty quantification at the cost of increased computational demand. To alleviate some of the computational burden associated with these conventional methodologies, supervised deep learning techniques- such as Convolutional Neural Networks- have been introduced (Das et al., 2019; Das & Mukerji, 2020; Wu et al., 2020). Once trained on a subset of seismic data and their corresponding properties, these models can be used to rapidly predict properties across the entire dataset. However, a significant limitation lies in the scarcity of labelled training data, as labels in geophysics typically come from well logs, which are expensive and sparsely available. To address the limitations of supervised learning in seismic inversion, researchers have begun incorporating physical constraints directly into the training process of neural networks. The emergence of physics-informed neural networks (PINNs) has proven effective for geophysical inversion, as they reduce reliance on large labelled datasets by embedding governing physical laws into the learning process (e.g., Biswas et al., 2019; Dhara & Sen, 2022; Vashisth & Mukerji, 2022; Liu et al., 2023; Sharma et al., 2024). This integration also ensures that the predicted outputs remain physically and geologically consistent.

While these physics-guided workflows have enabled substantial advancements, their limitations- primarily computational- become increasingly evident in the context of problems with complex forward modeling like full waveform inversion (FWI; Castellanos et al., 2015; He and Wang, 2021; Rasht-Behesht et al., 2022; Lin et al., 2023; Yang and Ma., 2023). FWI requires solving the wave equation iteratively and accurately, often using high-resolution velocity models and fine discretization. Embedding such detailed physics into a neural framework requires repeated forward and backward passes through large, non-linear PDE solvers during training, pushing the limits of conventional hardware. The challenge is even more pronounced in large-scale 3D models or time-lapse (4D) seismic datasets, where both data volume and model complexity increase substantially. Although scaling up classical resources might seem like a natural solution, the long-standing trend described by Moore's Law (Moore, 1965; Schaller, 1997)- the doubling of transistors on a chip approximately every two years- has plateaued (Theis and Wong, 2017; Shalf, 2020), marking the onset of diminishing returns as classical architectures reach their physical and architectural limits. These constraints highlight the need for a fundamentally different computing paradigm. In this context, quantum computing emerges as a compelling alternative, with the potential to accelerate large-scale geophysical workflows by leveraging quantum parallelism and the exponentially large state space of quantum systems.

Moreover, from an energy standpoint, quantum processors have been theoretically shown to offer significant reductions in energy consumption per computational task compared to classical high-performance computing for certain problem classes (Meier and Yamasaki, 2025), indicating that quantum approaches could become not only performance accelerators but also energy-efficient alternatives for large-scale scientific simulations.

Geophysics, with its reliance on high-dimensional data and complex computational tasks, has emerged as a promising yet relatively underexplored frontier for quantum computing and quantum machine learning (QML). There have been a few exploratory contributions to this field. Schade et al. (2024), demonstrated that 1D elastic wave equation could be solved exponentially faster using quantum computing compared to classical methods under certain conditions. Wright et al. (2024) presented quantum circuits tailored for simulating 1D acoustic wave equation. Additionally, quantum annealing- a specialized quantum computing paradigm designed for optimization- has also shown promise in geophysical applications. Dukalski et al. (2023) employed quantum annealers to estimate refraction residual statics for seismic data processing. Vashisth and Lessard (2024) and Vashisth et al. (2025) utilized quantum annealers to predict subsurface impedances from seismic data, showcasing the first application of quantum computing for seismic inversion. Further, Zaidenberg et al. (2021) and Sebastianelli et al. (2021, 2023) explored hybrid QML frameworks for image classification tasks in earth observation and remote sensing.

These advancements underscore the potential of quantum computing to address computational challenges in geosciences. Our study contributes to this ongoing effort by introducing a Hybrid Quantum Physics-Informed Neural Network (HQ-PINN), a framework that integrates quantum computing with physics-informed neural networks for the inversion of seismic data. Embedding physics into the quantum setting is particularly advantageous, as fully supervised training of quantum neural networks typically requires repeated quantum circuit evaluations over large datasets- an approach that can introduce substantial computational overhead and diminish potential quantum advantages. By incorporating physical constraints, the HQ-PINN framework circumvents reliance on labeled data and provides geologically and physically consistent solutions, offering a more efficient path for large-scale geophysical inversion. Note that our purpose here is not to compare with the current most efficient classical computing methods, but to lay the groundwork for applying quantum computing to prototypical geophysical applications.

Section 2 gives a general overview of quantum computing and quantum machine learning. Following sections describe the quantum node in a neural network, and its role in the proposed HQ-PINN framework for seismic inversion. The efficacy of HQ-PINNs is demonstrated through applications to synthetic seismic datasets, including both post-stack and pre-stack data. We also assess the performance of HQ-PINNs in estimating subsurface impedances from seismic data by varying quantum layer configurations, differentiation methods, and quantum device simulators. The HQ-PINN framework is then applied to the Sleipner dataset, showcasing its capability to handle real-world seismic data. We explore the impact of different types of qubit noise (depolarizing, thermal relaxation, dephasing, and readout) on the performance of HQ-PINN. Finally, we discuss the broader implications of leveraging HQ-PINNs for geophysical inversion and its potential to address overarching challenges in the field of computational geosciences.

## Quantum Computing and Quantum Machine Learning

Richard Feynman's seminal 1982 lecture "*Simulating Physics with Computers*", often considered the foundation of quantum computing, introduced the idea of using quantum systems to simulate complex quantum phenomena, recognizing the exponential cost of simulating such systems with classical computers (Feynman, 1982; Preskill, 2023). This concept was further advanced by David Deutsch, who raised the critical question of whether quantum computers could outperform classical systems in solving problems beyond quantum physics. He subsequently formalized the principles of universal quantum computation, laying the groundwork for practical exploration (Deutsch, 1985). The transformative moment, however, came with Peter Shor's groundbreaking algorithm for integer factorization (Shor, 1994), which provided concrete evidence that quantum computers could solve certain real-world problems exponentially faster than classical machines- capturing global scientific interest and propelling quantum computing into mainstream discourse.

At the core of a quantum computer lies the quantum bit, or qubit, which leverages fundamental quantum properties like superposition and entanglement to encode and process information in ways fundamentally distinct from classical bits (Nielsen and Chuang, 2010). While classical bits exist in one of two definite states, 0 or 1, qubits exist in a superposition of both states until measured, allowing quantum algorithms to explore multiple computational paths simultaneously through quantum interference. This enables certain types of parallel computations that have no classical counterpart. Additionally, quantum entanglement- a phenomenon where qubits exhibit strong correlations regardless of spatial separation- facilitates coordinated operations between qubits, enhancing computational efficiency. While entanglement does not enable faster-than-light communication, it plays a crucial role in quantum information processing, quantum teleportation, and error correction schemes necessary for scalable quantum computation. Together, these principles position quantum computers as potential game-changers, capable of outperforming classical systems in specialized problem domains. Researchers have increasingly sought to leverage these capabilities for diverse applications, including solving linear systems of equations (Harrow et al., 2009; Bravo-Prieto et al., 2023; Wang et al., 2024b), optimization problems (Perdomo-Ortiz et al., 2012; Farhi et al., 2014; Peruzzo et al., 2014; Neukart et al., 2017), and machine learning (Schuld et al., 2015).

Quantum Machine Learning (QML) represents a rapidly emerging synergy between quantum computing and traditional machine learning, aiming to augment the latter by leveraging the computational advantages of quantum systems. The implementation of QML varies based on whether its key components- data representation and computational processing- are realized using classical or quantum methodologies (Alchieri et al., 2021). However, the current era of Noisy Intermediate-Scale Quantum (NISQ) devices introduces significant hardware limitations, such as qubit decoherence, gate errors, and restricted scalability, which constrain the feasibility of fully quantum implementations for large-scale machine learning tasks (e.g., Preskill, 2018). To address these challenges, hybrid quantum-classical frameworks have emerged as a practical alternative. These frameworks integrate quantum and classical processors, leveraging their respective strengths. By offloading specific computational tasks to classical systems and reserving quantum processors for areas where they offer a distinct advantage, hybrid algorithms mitigate the limitations of NISQ devices. This approach enables the development of practical and scalable solutions to tackle complex computational challenges in the near term. The interdisciplinary impact of quantum computing continues to expand as it addresses computationally intensive challenges across diverse fields. A description of the foundations of quantum computing and quantum machine learning can be found in Appendix A. Appendix B and C describe commonly used single qubit gates and their operations on multi-qubit circuits, respectively.

## Quantum Nodes for Hybrid Quantum Neural Networks (HQNNs)

Hybrid Quantum Neural Networks (HQNNs) integrate quantum circuits into classical neural network architectures to harness quantum parallelism (superposition and entanglement) while maintaining compatibility with classical optimization methods. At the heart of an HQNN lies the quantum node (QNode), which serves as a differentiable quantum layer within a hybrid model. It typically comprises three key components: i) a feature map for encoding classical input data into quantum states, ii) a parameterized quantum ansatz for transformation, and iii) measurement operations to extract classical information from the resulting quantum state.

Classical data can be embedded into quantum states using encoding schemes such as angle embedding or amplitude embedding (Schuld and Petruccione, 2021). In amplitude embedding, which is employed in this study, the input vector is normalized and directly mapped to the amplitudes of quantum state. This enables efficient representation of high-dimensional classical data (e.g., seismic data) within the quantum Hilbert space.

The ansatz is a parameterized quantum circuit that transforms the encoded quantum state to explore the solution space. In this work, we employ the basic entangler layers ansatz, which consists of parameterized single-qubit rotation gates followed by entangling CNOT gates arranged in a closed-chain topology (see Appendix A for a mathematical description of the CNOT gate). This design is considered hardware-efficient (Kandala et al., 2017; Sim et al., 2019), and is well-suited for training in hybrid quantum neural networks. A comparison with the strongly entangling layers ansatz, which extends the basic entangler design to achieve deeper entanglement through range-based two-qubit interactions, is discussed in the *Discussion* section.

Following the transformation by the ansatz, measurements are performed on the quantum state- typically as expectation values of Pauli operators- to extract real-valued outputs. These outputs can either be used directly in the loss function or passed as inputs to subsequent classical neural network layers.

The quantum node is implemented using the PennyLane library (Bergholm et al., 2018), which supports automatic differentiation via different methods such as the adjoint method, parameter-shift rule, and Simultaneous Perturbation Stochastic Approximation (SPSA). This enables seamless integration of quantum circuits into classical machine learning pipelines, facilitating the training of hybrid models such as the HQ-PINN. Brief descriptions of the different auto differentiation methods mentioned above can be found in Appendix D.

## Hybrid Quantum Physics-Informed Neural Networks (HQ-PINNs) for Seismic Inversion

The HQ-PINN architecture (Fig. 1) is structured as an encoder-decoder framework that integrates hybrid quantum-classical computations to perform seismic inversion. The encoder is a HQNN that takes seismic data as an input and outputs a latent representation of the elastic parameters. The input layer is a quantum node that takes seismic data and employs the amplitude embedding feature map to leverage quantum superposition of $\boldsymbol{n}$ qubits, enabling the encoding of data of size $\mathbf{2}^{\boldsymbol{n}}$. When the size of seismic data does not match $\mathbf{2}^{\boldsymbol{n}}$, it is padded with zeros to conform to the next nearest power of 2 before normalization for amplitude embedding. We employ the basic entangler ansatz, consisting of two layers (Fig. 2). Each layer comprises parameterized single-qubit rotational gates, followed by a network of CNOT gates arranged in a closed-chain configuration. This configuration ensures that every qubit interacts with its neighbour, forming a continuous loop, making this structure well-suited for quantum neural networks. The quantum measurement yields the expectation values of the Pauli-Z operator (Fig. 2), which corresponds to measurements in the computational basis. The expectation values are subsequently fed into a single classical fully connected layer, where the number of output nodes corresponds to the total dimensionality of the elastic parameters to be estimated. This layer utilizes a sigmoid activation function to scale the outputs between 0 and 1, facilitating their rescaling within the known ranges of the elastic parameters before they are passed as input to the decoder.

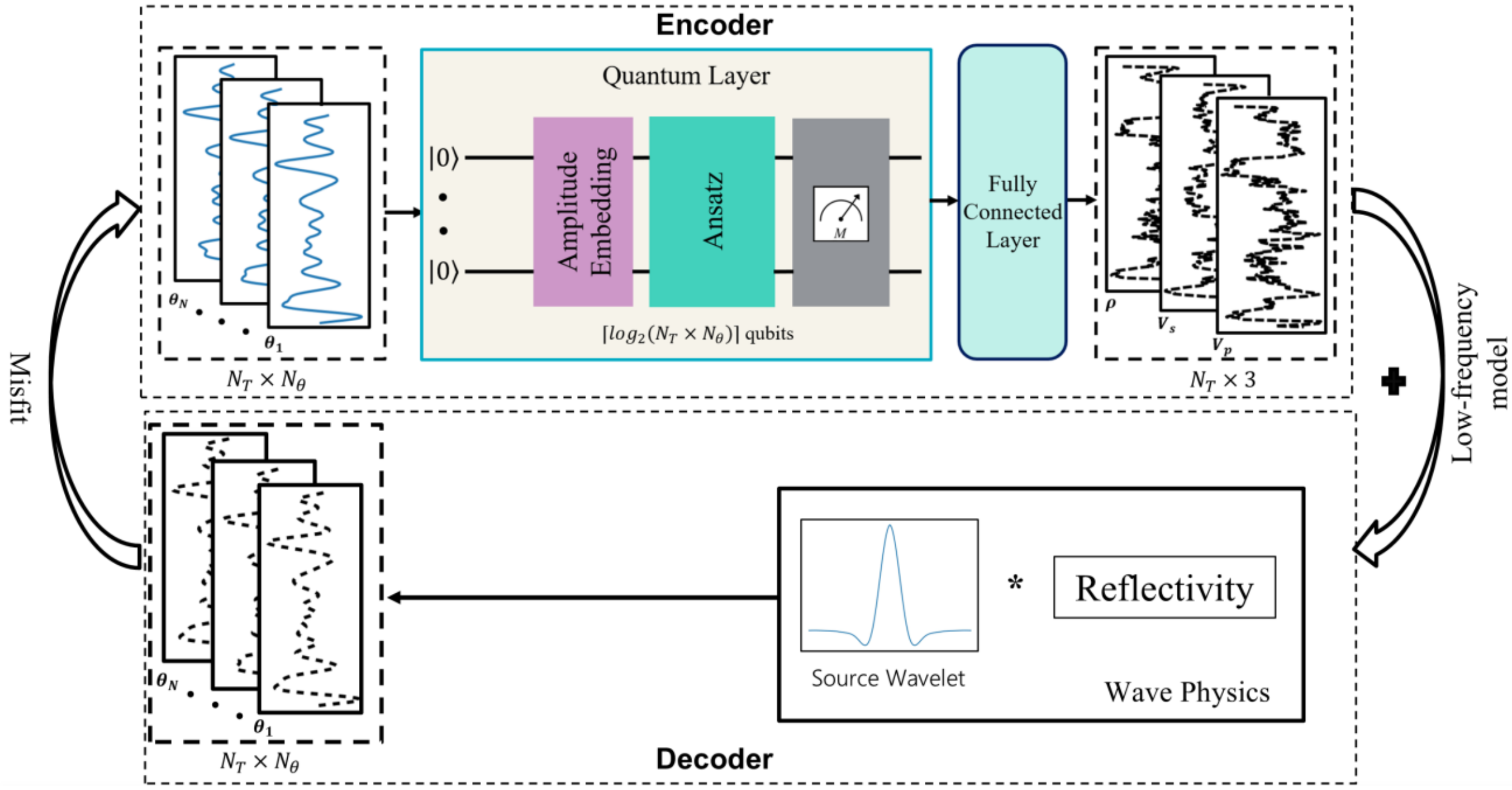

**Figure 1**. The HQ-PINN framework for estimating elastic parameters from input seismic data. $N_T$ represents the number of time samples, and $N_\theta$ denotes the number of angle gathers. The encoder is an HQNN model that predicts the elastic parameters, which are then passed to the decoder to generate the corresponding angle gathers based on geophysical relationships. The HQ-PINN model learns by minimizing the misfit between the input and predicted seismic data from the decoder.

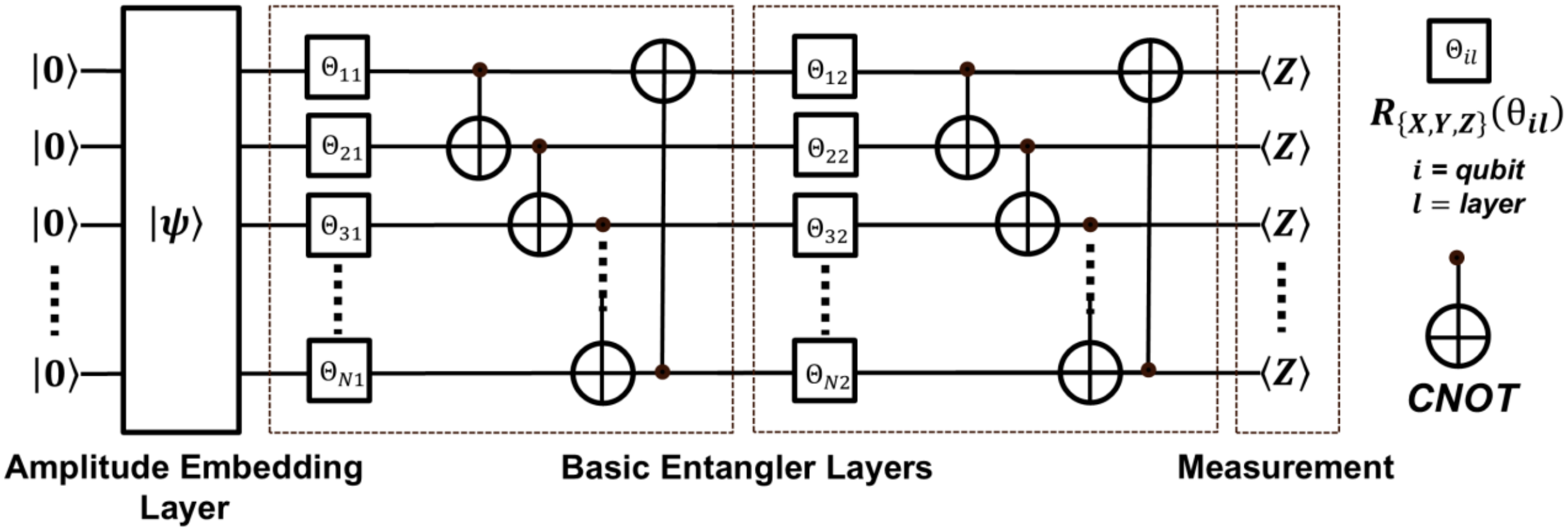

**Figure 2**. The quantum node/layer utilized in the encoder of the HQ-PINN framework for N-qubits. It employs an amplitude embedding feature map and basic entangler layers ansatz, with measurements performed in the computational basis.

The decoder block utilizes the elastic parameters predicted by the encoder to generate the corresponding pre-stack and post-stack seismic data using Aki-Richards relations (Aki and Richards, 2002) for the reflection coefficients. The loss function consists of seismic misfit and a regularization term. The seismic misfit is calculated as the root mean squared error (RMSE) between the true (observed) and predicted seismic data from the decoder. The regularization term penalizes deviations of the predicted elastic parameters from a background or low-frequency trend. In seismic inversion, a low-frequency model is often incorporated (in this case, as a regularization term) to compensate for the lack of low-frequency information in the seismic data. This low-frequency model helps stabilize the inversion process and ensures that the solution remains physically meaningful. After computing the loss, TensorFlow's/ PyTorch's automatic differentiation is used to update the model parameters and backpropagate the misfit through the decoder into the encoder, ensuring efficient end-to-end optimization. A key step in this process involves computing gradients within the quantum circuit. Unless otherwise specified, we employ the adjoint differentiation method to calculate exact derivatives on the *lightning.qubit* simulator. Furthermore, we use the Adam optimizer (Kingma and Ba, 2014) with a learning rate of 0.1 to update the model parameters during training.

In this study, HQ-PINN operates in an unsupervised setting, leveraging geophysical relationships to learn the underlying physics and estimate P- and S-impedances from the input seismic data. The HQ-PINN framework is highly flexible and can be readily adapted to a supervised setting when sufficient labelled data are available for training the network.

## Synthetic Post-stack Seismic Inversion

For the synthetic post-stack seismic case study, we employed the same dataset as that used in Vashisth and Mukerji (2022) but focused on estimating acoustic impedances from seismic traces instead of porosities. Each seismic trace comprises 246 time samples; to facilitate amplitude embedding using 8 qubits, the traces are padded with zero amplitudes to a length of 256. Consequently, the quantum layer is configured as an 8-qubit quantum node with an amplitude embedding feature map and a basic entangler layers ansatz. We experimented with various configurations of the basic entangler layers ansatz, incorporating $\boldsymbol{R_X}$, $\boldsymbol{R_Y}$ and $\boldsymbol{R_Z}$ rotation gates (see appendix B for descriptions of rotation gates). Additionally, we analyzed a configuration without any ansatz, where the quantum layer includes only the feature map and no trainable parameters in the quantum circuit. In this scenario, the trainable parameters solely reside in the fully connected layer. The results, including the RMS misfit between the true and estimated acoustic impedance profiles (encoder), the true and predicted seismic traces (decoder), and the training times of the HQ-PINN model, are summarized in *Table 1*. It should be noted that the adjoint differentiation method and *lightning.qubit* simulator is used for optimizing the quantum layer. Similar results are observed across the basic entangler layer configurations utilizing $\boldsymbol{R_X}, \boldsymbol{R_Y}$ and $\boldsymbol{R_Z}$ rotation gates. The estimated acoustic impedance profile and seismic trace, from the trained HQ-PINN model incorporating the $\boldsymbol{R_X}$ rotation gate, are presented in Fig. 3 (example 1). Notably, the no-ansatz configuration also demonstrated the ability to predict accurate impedance profiles, albeit with slightly higher misfits and faster training times. These findings underscore the effectiveness of the embedding layer and its capability to encode seismic data in the quantum Hilbert space, enabling accurate seismic inversion even in the absence of an ansatz (i.e., no trainable parameters in the quantum layer), positioning the QNode as a promising candidate for an efficient dimensionality reduction tool in solving high-dimensional inversion problems.

In addition to exploring different ansatz structures, we also investigated various differentiation methods for computing gradients within the quantum node. For this experiment, we fixed the basic entangler layers ansatz with $\boldsymbol{R_X}$ rotation gates and compared the performance of four differentiation methods- adjoint, parameter-shift, finite-difference, and SPSA- in terms of accuracy and computational speed. *Table 2* lists the RMS misfit between the true and predicted acoustic impedance profiles and seismic traces, along with the corresponding HQ-PINN training times. The impedance and seismic misfits are nearly identical across all differentiation methods, with SPSA yielding slightly higher misfit values. However, notable differences arise in computational efficiency. The adjoint differentiation method is the fastest (10.49 s), followed by SPSA (28.44 s), finite-difference (92.03 s), and parameter-shift (150.55 s) approaches.

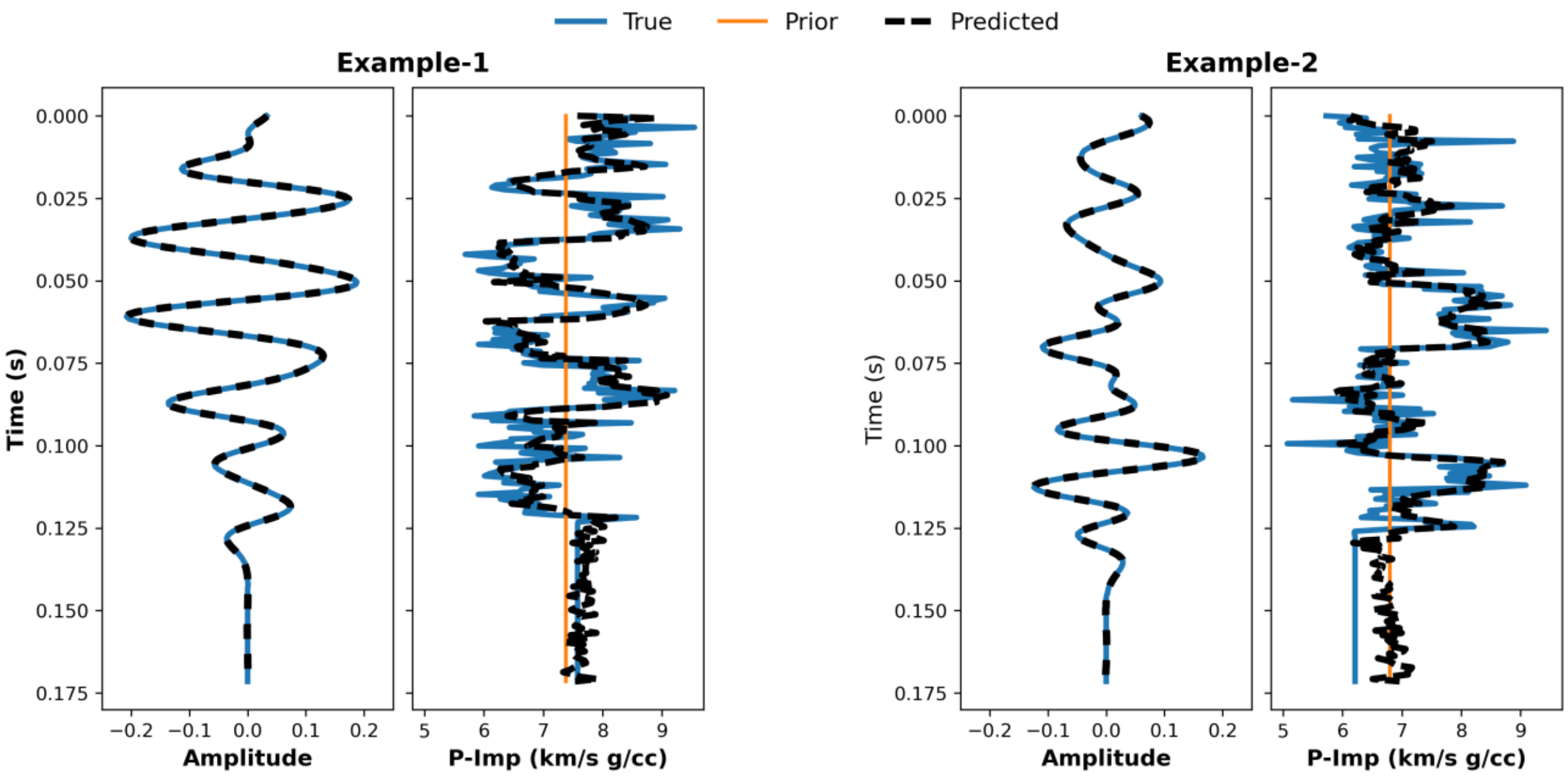

**Figure 3**. HQ-PINN results for synthetic post-stack seismic examples. True seismic and acoustic impedance models are shown in blue, HQ-PINN predicted models in black, and the low-frequency prior is depicted in orange.

**Table 1.** RMS misfits in the estimated seismic and acoustic impedance profiles, along with the training time for HQ-PINNs with the basic entangler layers ansatz with $\{R_X, R_Y$ and $R_Z\}$ rotation gates, and the case without an ansatz for post-stack inversion.

| Ansatz Configuration | P-imp misfit (km s⁻¹ g cm⁻³) | Seismic misfit | Training time (s) |
|---|---|---|---|
| $\boldsymbol{R_X}$ | 0.3983 | 0.000563 | 10.49 |
| $\boldsymbol{R_Y}$ | 0.3943 | 0.000553 | 10.82 |
| $\boldsymbol{R_Z}$ | 0.3970 | 0.000564 | 10.85 |
| **No Ansatz** | 0.4133 | 0.000801 | 6.63 |

**Table 2.** RMS misfits in the estimated seismic and acoustic impedance profiles, along with the training time for HQ-PINNs with different differentiation methods in the quantum layer for post-stack inversion.

| Differentiation method | P-imp misfit (km s⁻¹ g cm⁻³) | Seismic misfit | Training time (s) |
|---|---|---|---|
| **Adjoint** | 0.3983 | 0.000563 | 10.49 |
| **Parameter-shift** | 0.3983 | 0.000563 | 150.55 |
| **Finite-difference** | 0.3982 | 0.000566 | 92.03 |
| **SPSA** | 0.4099 | 0.001030 | 28.44 |

The adjoint differentiation method achieves the fastest training time because it computes exact gradients through reverse-mode automatic differentiation, which efficiently backpropagates errors through all model parameters in a single backward pass. Its computational cost scales roughly with the cost of one forward model evaluation, making it particularly efficient for complex problems. SPSA ranks second in speed because it estimates the gradient using only two forward evaluations- regardless of the number of parameters- by perturbing all parameters simultaneously with random noise. Although this stochastic approximation introduces higher gradient variance (leading to slightly higher misfits), its dimension-independent cost and robustness to measurement noise make it attractive for problems with a large number of parameters or hardware implementations where exact adjoint gradients are unavailable or memory-intensive. Finite-difference methods, though theoretically accurate, are slower due to requiring two model evaluations per parameter (forward and backward perturbations), resulting in a computational cost that scales linearly with the number of parameters and can be prone to numerical errors caused by step-size sensitivity. Parameter-shift, while highly accurate for quantum systems due to its exact gradient computations, is the most computationally expensive, as it requires two circuit evaluations per quantum parameter to compute exact analytical gradients via expectation-value shifts. For complex problems, the trade-off between computational cost and accuracy becomes more significant, and selecting or adapting differentiation methods to the problem's characteristics is essential.

We applied the HQ-PINN model with a basic entangler layers ansatz utilizing $\boldsymbol{R_x}$ gates and the adjoint differentiation method to invert another post-stack seismic trace (Example 2, Fig. 3). The predicted acoustic impedance profile and seismic trace demonstrate a close match with their true counterparts, as illustrated in Fig. 3. The RMS misfits for the acoustic impedance and seismic data are 0.4692 km $s^{-1}$ g $cm^{-3}$ and 0.000677, respectively, with the training time similar to that of Example 1.

## Synthetic Pre-stack Seismic Inversion

To further assess the efficacy of HQ-PINNs, we extend their application to the inversion of pre-stack seismic data. For this study, we utilized the P- and S- impedance dataset provided by Das and Mukerji (2020) and generated angle gathers at incident angles of 5°, 15°, and 25° using a 40 Hz Ricker wavelet. We tested the HQ-PINN framework on two examples, as illustrated in Fig. 4. For these examples, the quantum layer of the encoder is configured with a 10-qubit quantum node, utilizing an amplitude embedding feature map and a basic entangler layer ansatz with $\boldsymbol{R_X}$ gates. The HQ-PINN model is trained for 500 epochs, with the entire training process completed in approximately 14.6 s. The RMS misfits between the true and predicted P- and S- impedance profiles, as well as the true and predicted seismic data across all angle gathers, are summarized in *Table 3*. The predicted P- and S-impedances (encoder), along with their seismic angle gathers (decoder), closely match the true data (Fig. 4), as indicated by the low RMSE values. These results underscore the capability of HQ-PINNs to efficiently estimate impedance profiles from pre-stack seismic data.

**Table 3.** RMS misfits in the HQ-PINN estimated seismic, acoustic and shear impedance profiles for pre-stack inversion examples.

| Misfit | Example 1 | Example 2 |
| --- | --- | --- |
| Seismic [$5^\circ$] | 0.00096 | 0.00085 |
| Seismic [$15^\circ$] | 0.00101 | 0.00089 |
| Seismic [$25^\circ$] | 0.00117 | 0.00104 |
| P-imp (km $s^{-1}$ g $cm^{-3}$) | 0.4722 | 0.4216 |
| S-imp (km $s^{-1}$ g $cm^{-3}$) | 0.2398 | 0.2395 |

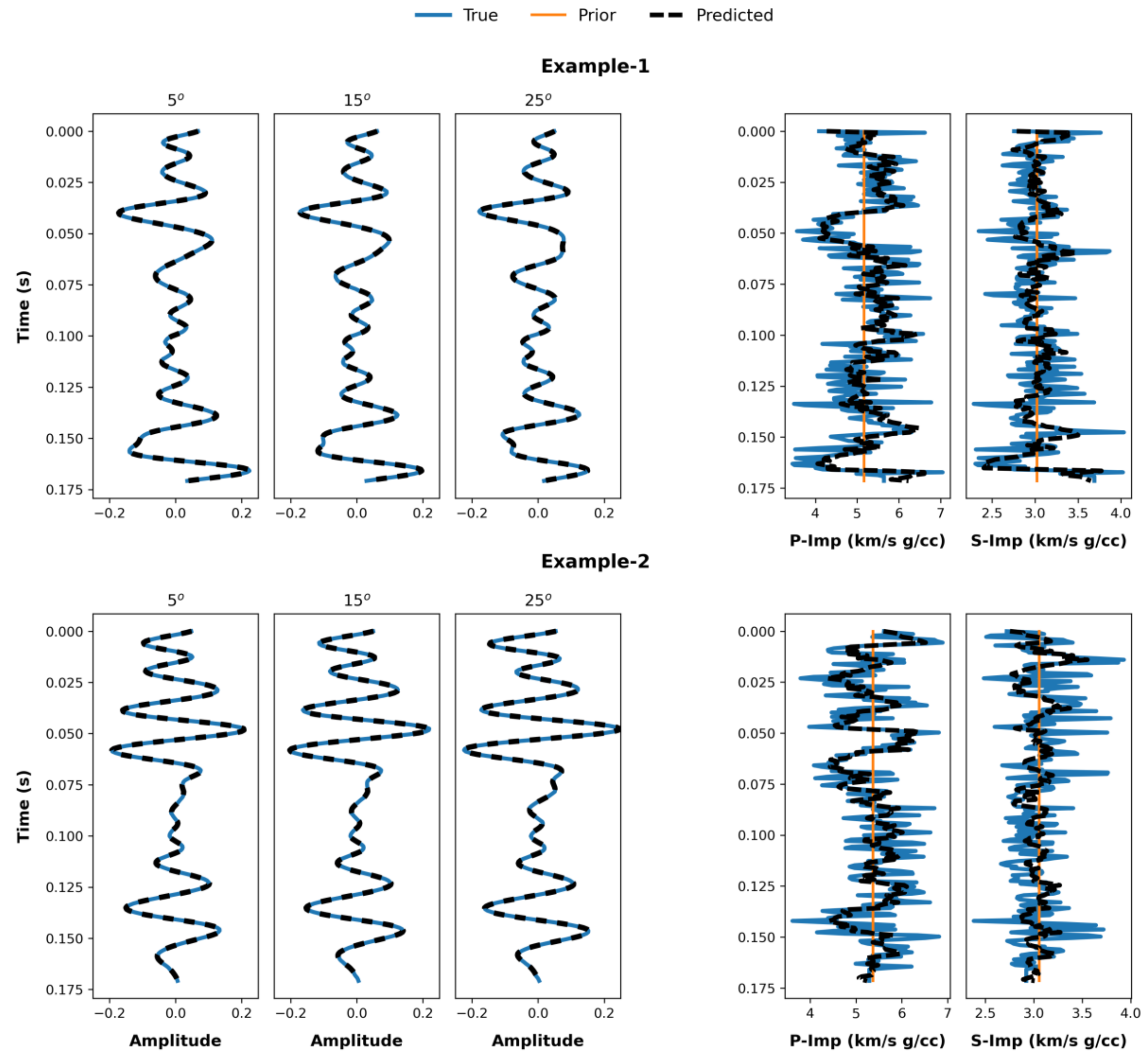

**Figure 4**. HQ-PINN results for synthetic pre-stack seismic examples with angle gathers of 5°, 15°, and 25°. True seismic angle gathers, acoustic and shear impedances are shown in blue, HQ-PINN predicted models in black, and the low-frequency prior is depicted in orange.

## Performance on the Sleipner Dataset

To evaluate the real-world applicability of the HQ-PINN model, we applied it to the well-known Sleipner dataset, derived from the Sleipner $CO_2$ storage project in the North Sea- one of the world's first large-scale $CO_2$ sequestration initiatives (Torp and Gale, 2004). This dataset includes seismic data acquired from the Utsira Sand formation, a highly porous sandstone reservoir overlain by impermeable caprock, making it ideal for $CO_2$ storage. The dataset captures the temporal evolution of $CO_2$ injection into the reservoir, providing valuable insights into the spatial and volumetric distribution of $CO_2$ over time. Romero et al. (2023) implemented a joint inversion-segmentation approach on the Sleipner 4D post-stack seismic dataset. We used the same dataset and wavelets to evaluate the HQ-PINN framework, and constructed low-frequency prior model by integrating seismic horizons with available well data. Specifically, we focus on 2D subsections (200 × 200) around the Utsira formation for the baseline (1994) and monitor (2001) surveys, where $CO_2$ injection had occurred.

To calibrate the HQ-PINN workflow, we first inverted the inline seismic subsection from the baseline survey passing through the well location (Baseline-1). A 16-qubit quantum node is employed, utilizing an amplitude embedding feature map and a basic entangler layers ansatz with $\boldsymbol{R_x}$ gates. The HQ-PINN model is trained for 200 epochs, and the predicted acoustic impedances (encoder output) are presented in Fig. 5 (Baseline-1). The inversion minimizes a composite objective function comprising data misfit and Tikhonov regularization terms that enforce consistency with a low-frequency prior and promote spatial smoothness via a Laplacian operator. For comparison, we also performed conventional least-squares inversion with Tikhonov regularization (LS-TR, Romero et al., 2023). The well data and the acoustic impedances derived from both the HQ-PINN model and the LS-TR method are displayed in Fig. 6. At the well location, the estimated acoustic impedances show close alignment with the impedance log data, with RMSE of 0.551 km $s^{-1}$ g $cm^{-3}$ for the HQ-PINN model and 0.605 km $s^{-1}$ g $cm^{-3}$ for the LS-TR method. Furthermore, the observed and predicted seismic sections (decoder) based on the estimated acoustic impedance model (encoder) exhibit strong agreement, achieving an RMSE of 0.0944, as shown in Fig. 7. Additionally, we inverted an inline seismic subsection near the $CO_2$ injection site from the baseline survey using the same HQ-PINN configuration. The predicted acoustic impedance model and seismic section for this case (Baseline-2) are shown in Figs 5 and 7, respectively. The RMSE between the true and predicted seismic sections for the Baseline-2 survey is 0.0919. The training time for both baseline surveys is approximately 109 seconds using the *lightning.qubit* simulator device.

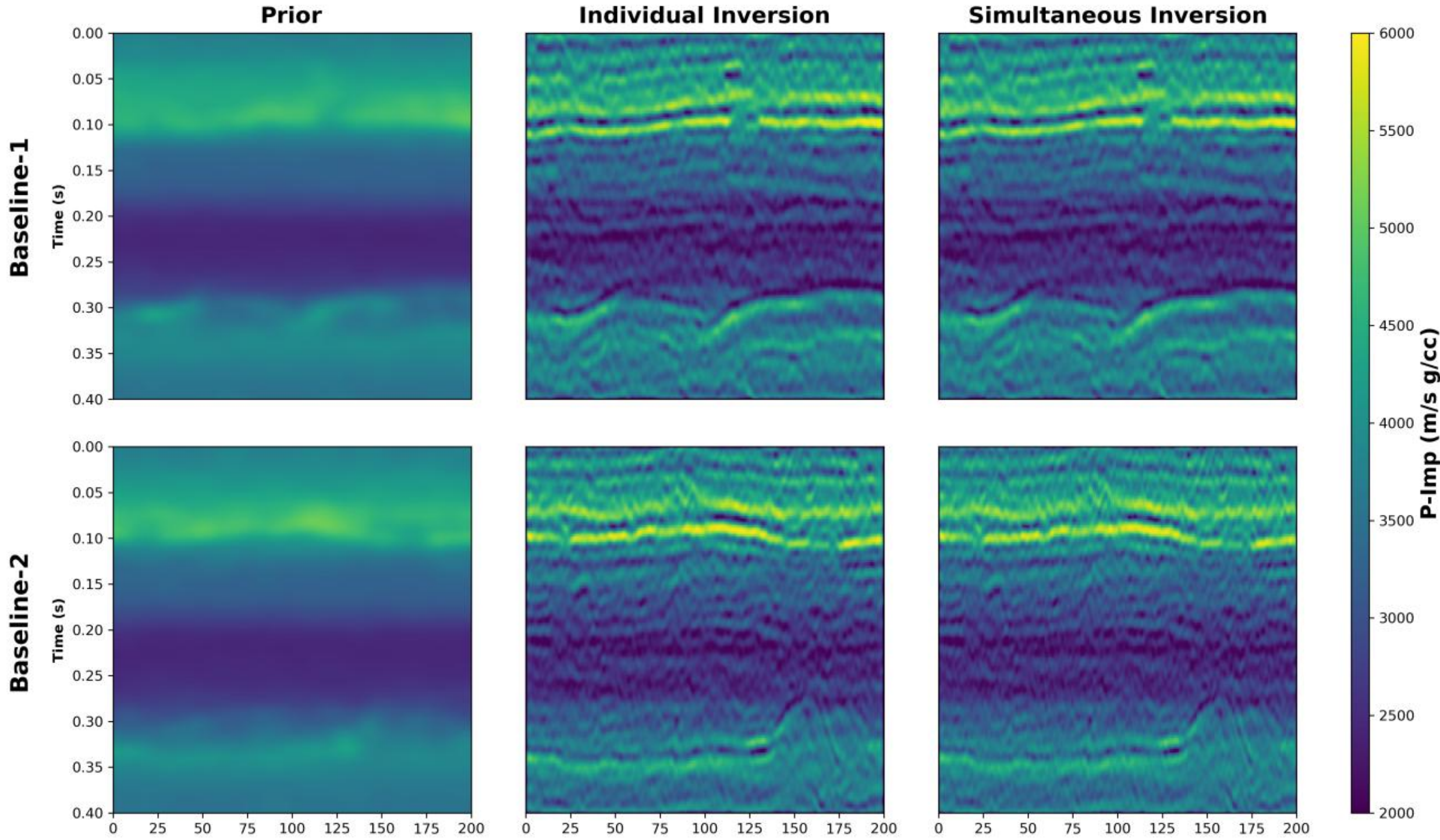

**Figure 5**. HQ-PINN predicted acoustic impedance models for both individual and simultaneous inversion of baseline seismic surveys: Baseline-1, passing through the well location, and Baseline-2, near the $CO_2$ injection site. The results are demonstrated alongside the low-frequency prior used in this study.

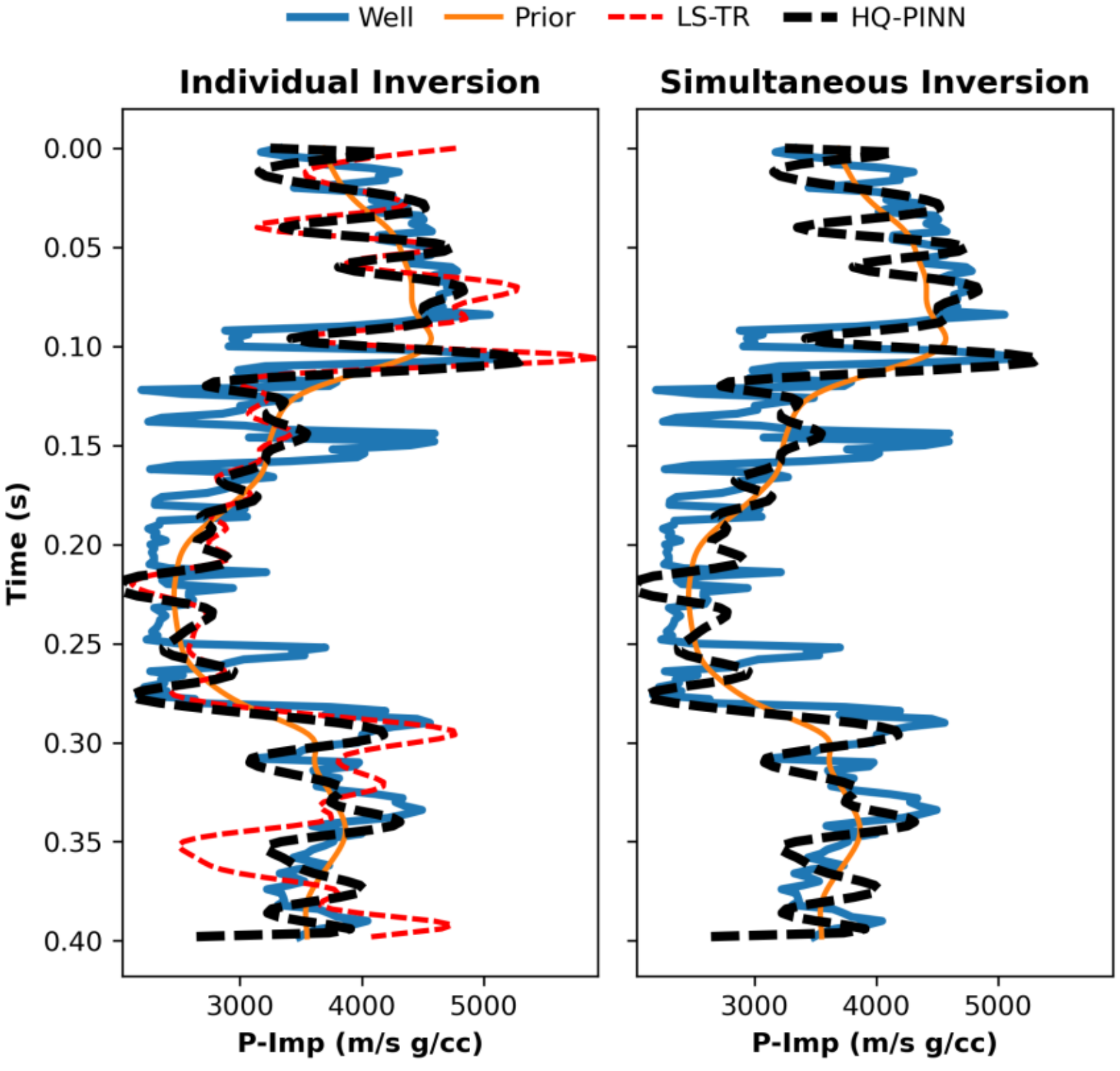

**Figure 6**. HQ-PINN predicted acoustic impedances for both individual and simultaneous inversion of the Sleipner baseline seismic dataset (Baseline-1) at the well location. The predictions are compared to the impedance log, the low-frequency prior model, and the results from LS-TR inversion for the individual inversion case.

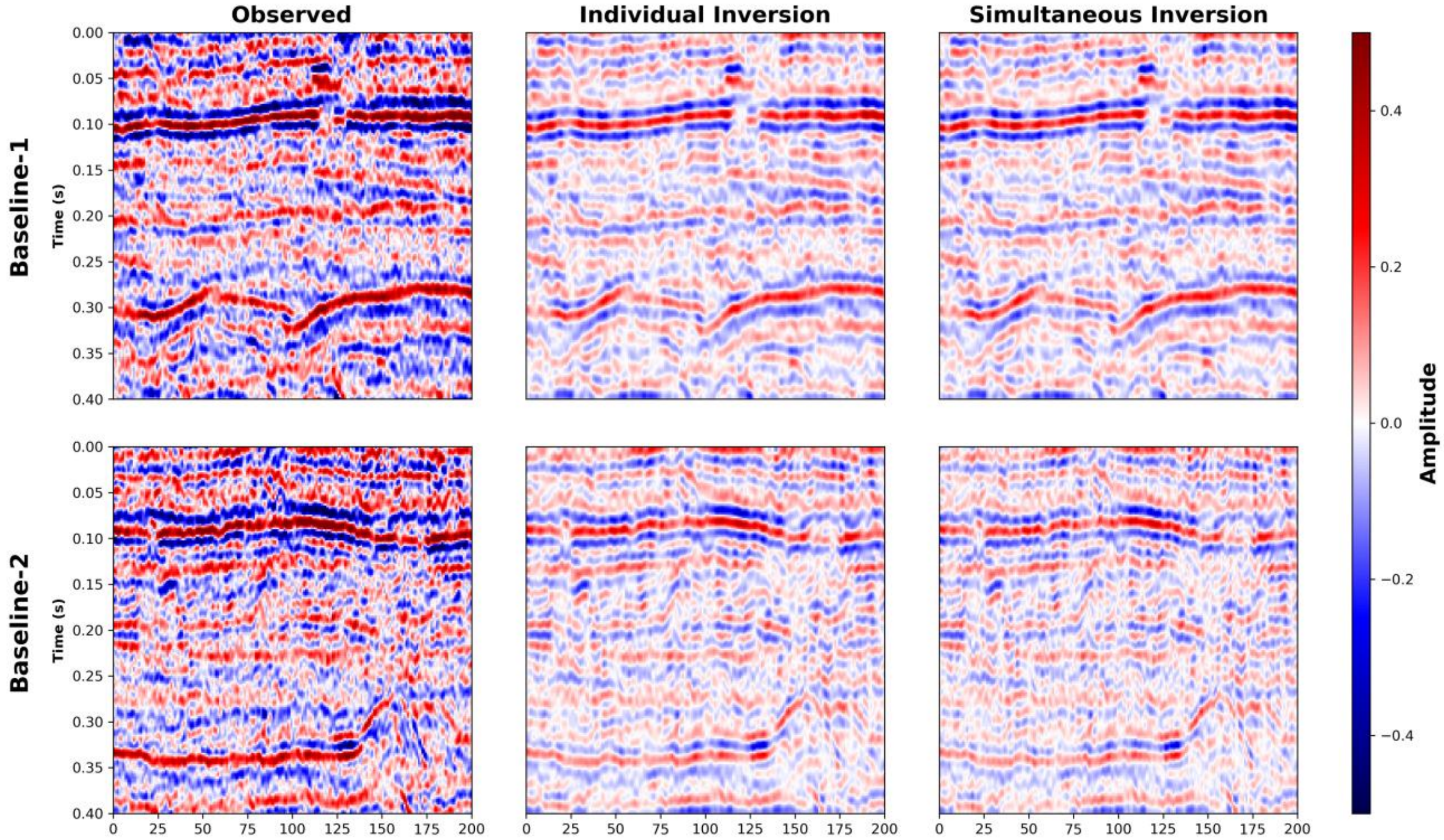

**Figure 7**. Observed and predicted seismic sections from the HQ-PINN decoder for both individual and simultaneous inversion of baseline seismic surveys: Baseline-1, passing through the well location, and Baseline-2, near the $CO_2$ injection site.

To demonstrate the scalability and computational efficiency of integrating quantum technology into classical computing workflows, we also inverted both the seismic sections (Baseline-1 and Baseline-2) simultaneously. Notably, this simultaneous inversion required only one additional logical qubit to encode and invert both 2D seismic sections. By increasing the number of qubits in the quantum layer from 16 to 17, we input both seismic sections into the HQ-PINN framework and estimate acoustic impedance models similar to those predicted with separate individual inversions. Figs 5 and 7 illustrate the estimated acoustic impedance models and seismic sections for both baseline surveys using the HQ-PINN framework. Fig. 6 shows that the estimated acoustic impedances at the well location closely aligns with the individual inversion prediction and shows good agreement with the impedance log, achieving an RMSE of 0.551 km $s^{-1}$ g $cm^{-3}$.

For all previous case studies, we used the *lightning.qubit* simulator as the quantum device. This is because for quantum circuits with relatively low qubit and gate counts, the overhead associated with GPU memory transfer and kernel initialization outweighs the computational benefits of parallelization. In such scenarios, the highly optimized CPU-based implementation in *lightning.qubit* achieves faster execution times. However, for circuits involving higher qubit counts and dense gate operations, the parallelism inherent in GPU architectures significantly accelerates computations. This is evident in the Sleipner field case study, where the simultaneous inversion required 17 qubits. Utilizing the *lightning.gpu* simulator led to a substantial reduction in training time- from 203.03 seconds to 45.78 seconds (a 77.5% decrease)- while maintaining identical accuracy in results. We utilized a single GPU, and further reductions in training time may be achieved by leveraging multiple GPUs in parallel. Furthermore, the SPSA differentiation method on the *lightning.gpu* simulator achieved similar results with a slightly reduced training time of 42.13 seconds. Pushing the limits further, we removed the ansatz from the quantum layer, utilizing the quantum layer solely as a dimensionality reduction tool, with learnable parameters confined to the output classical fully connected layer. Remarkably, this configuration also yielded comparable and accurate results, with training times of 64.36 seconds on the *lightning.qubit* device and 20.99 seconds on the *lightning.gpu* device, further demonstrating the efficiency and efficacy of the HQ-PINN framework. All results for the Sleipner simultaneous inversion case study are summarized in *Table 4*.

We further extend our analysis to $CO_2$ monitoring near the injection site. The acoustic impedance model estimated by the HQ-PINN framework for the Baseline-2 survey (1994) is smoothed and utilized as the initial/prior model for the monitor survey (2001). The resulting estimated acoustic impedance model and seismic section for the monitor survey are presented in Fig. 8, showing an RMSE of 0.1715 between the observed and predicted seismic data. Notably, the HQ-PINN framework successfully delineates the $CO_2$ plume in both the estimated seismic section and acoustic impedance model.

**Table 4.** RMS misfits in the HQ-PINN estimated seismic sections for the simultaneous inversion of two baseline seismic surveys, along with the training time across various quantum devices (simulators), ansatz configurations and differentiation methods.

| **Device** | ***lightning.qubit*** | ***lightning.gpu*** | ***lightning.gpu*** | ***lightning.qubit*** | ***lightning.gpu*** |
|---|---|---|---|---|---|
| **Ansatz** | **B.E. with $R_X$** | **B.E. with $R_X$** | **B.E. with $R_X$** | **No Ansatz** | **No Ansatz** |
| **Diff. method** | **Adjoint** | **Adjoint** | **SPSA** | **-** | **-** |
| **Baseline-1** | 0.0944 | 0.0944 | 0.0944 | 0.0943 | 0.0943 |
| **Baseline-2** | 0.0955 | 0.0955 | 0.0955 | 0.0952 | 0.0952 |
| **Time (s)** | 203.03 | 45.78 | 42.13 | 64.36 | 20.99 |

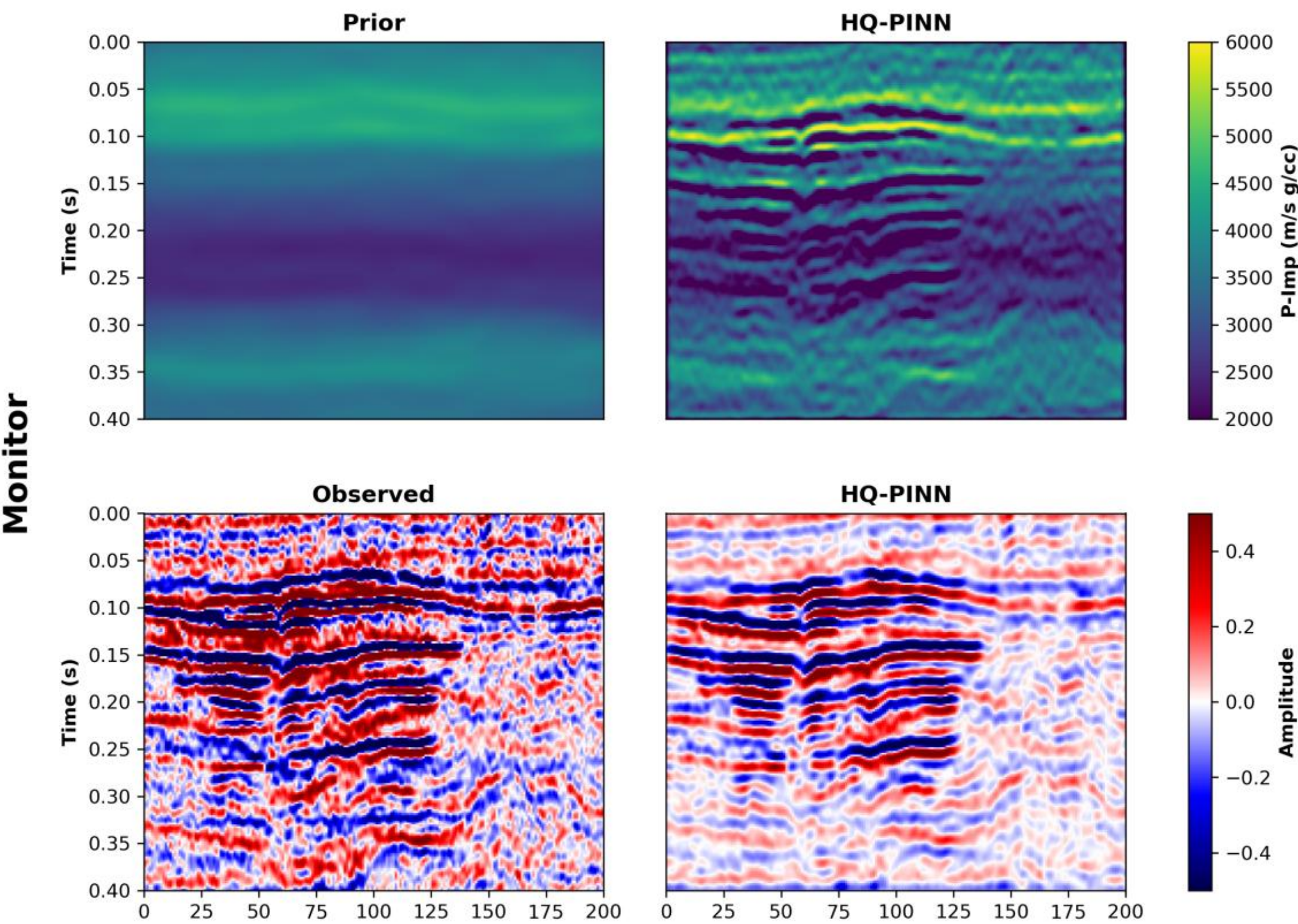

**Figure 8**. HQ-PINN predicted seismic and acoustic impedance model for the individual inversion of monitor seismic survey (observed) near the $CO_2$ injection site.

## Impact of Noisy Simulators mimicking Quantum Hardware

Quantum computing today operates in the NISQ era, where quantum processors contain limited numbers of qubits that are prone to decoherence, gate infidelity, and unwanted environmental coupling. As a result, all current quantum devices suffer from imperfect control, leading to noise in quantum operations and measurements. To realistically evaluate how the proposed HQ-PINN framework would perform on real hardware, it is therefore essential to include hardware-level noise within simulations. To achieve this, we employed the PennyLane-Qiskit plugin in conjunction with a Qiskit fake backend. Specifically, we used the *GenericBackendV2* configuration, which allows users to create customizable mock quantum devices for local testing and performance benchmarking. The backend was defined by specifying the number of qubits (eight for our post-stack inversion problem) and the basis gate set on which noise channels were applied. In our case, the noisy gate set included the identity (**I**), single-qubit rotation gates ($\boldsymbol{R_X}$, $\boldsymbol{R_Y}$), the two-qubit controlled-NOT (CNOT), and measurement operations. Additional hardware

characteristics- such as coherence times, resonance frequencies, and gate fidelities- were sampled from realistic ranges extracted from historical IBM Quantum backend data.

Three principal sources of noise were modelled. **Depolarizing noise** represents gate infidelity, where each gate acts correctly with probability $1 - p$, and with probability $p$ the qubit state is replaced by the maximally mixed state $I/2$, effectively randomizing its information. **Thermal relaxation** and **dephasing** capture finite qubit lifetimes, characterized respectively by the $T_1$ and $T_2$ time constants. Relaxation models the spontaneous decay of the excited state $|\mathbf{1}\rangle$ to the ground state $|\mathbf{0}\rangle$, while dephasing models the randomization of phase in quantum superpositions, which gradually drives the system toward classicality (Georgopoulos et al., 2021). Finally, **measurement (readout) errors** were introduced to account for misclassification of qubit outcomes, parameterized by $P(0|1)$ and $P(1|0)$- the probabilities of recording an incorrect classical result during measurement.

For our 8-qubit synthetic post-stack inversion case, we configured a *GenericBackendV2* instance using these specifications. The resulting device-level parameters are summarized in Table 5, including the coherence times $(T_1, T_2)$ and qubit resonance frequencies, while Table 6 lists the depolarizing probabilities and operation times for each gate. The results show that single-qubit operations ($\boldsymbol{R_X}$, $\boldsymbol{R_Y}$, $\boldsymbol{I}$) maintain high fidelities and low variance in depolarizing probability across all qubits, whereas more complex two-qubit CNOT gates and measurements exhibit substantially higher variance- up to roughly 60 %. Operation times display a similar trend. These findings are consistent with the known performance characteristics of superconducting-qubit architectures (Kjaergaard et al., 2020).

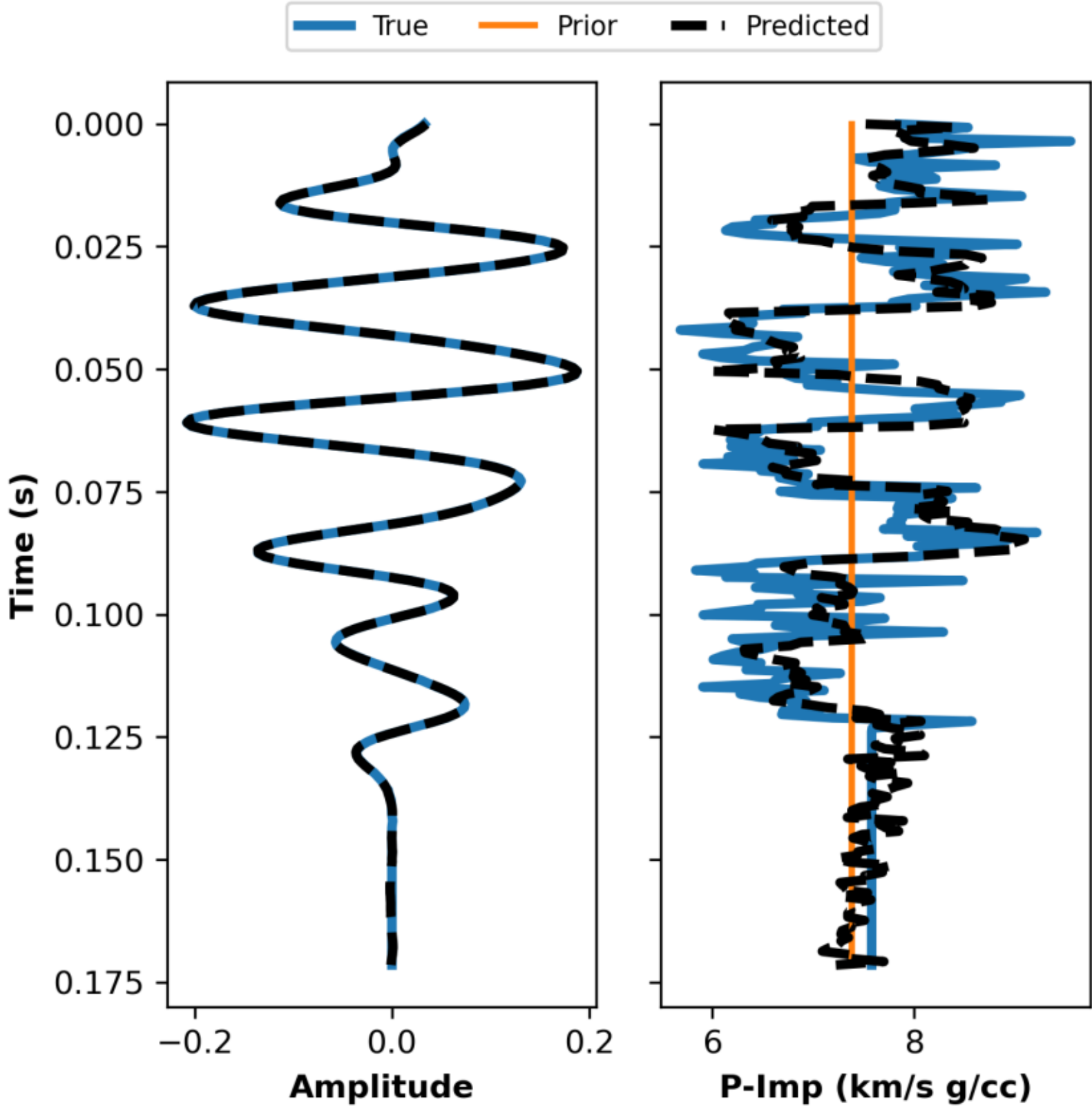

**Figure 9**. HQ-PINN results from noisy simulator (fake backend) for synthetic post-stack seismic example- 1. True seismic and acoustic impedance models are shown in blue, HQ-PINN predicted models in black, and the low-frequency prior is depicted in orange.

**Table 5.** Qubit-level properties (coherence times and resonance frequencies) for the HQ-PINN quantum node.

| Qubit | T1 (Relaxation time) (µs) | T2 (Dephasing time) (µs) | Frequency (GHz) |
|---|---|---|---|
| **1** | 177 | 144 | 5.4293 |
| **2** | 170 | 109 | 5.4878 |
| **3** | 176 | 179 | 5.0641 |
| **4** | 145 | 137 | 5.4634 |

| **5** | 164 | 182 | 5.2217 |
|---|---|---|---|
| **6** | 123 | 155 | 5.0319 |
| **7** | 183 | 163 | 5.3790 |
| **8** | 135 | 197 | 5.4466 |

**Table 6.** Gate-level noise parameters (depolarizing probabilities and operation times) for the HQ-PINN quantum node.

| Operation | Depolarizing Probability | Operation time (ns) |
|---|---|---|
| **CNOT** | 0.001177 | 779.66 |
| **$R_X$** | 0.000098 | 43.07 |
| **$R_Y$** | 0.000096 | 43.73 |
| **Measure** | 0.00312 | 1317.52 |
| **Identity** | 0.000092 | 53.28 |

We retrained the same HQ-PINN model previously used on the noiseless simulator using the noisy backend, reducing the learning rate from 0.1 to 0.05 to ensure stable convergence under noise-induced fluctuations. The SPSA differentiation method was employed due to its robustness against stochastic gradient noise and computational efficiency under hardware-like conditions. A key computational bottleneck arises from the use of mixed-state simulators. When noise is introduced, quantum states evolve from pure states to mixed states represented by density matrices rather than state vectors. Simulating such systems requires tracking $2^{2n}$ complex amplitudes for an $n$-qubit register- an exponential overhead compared to the $2^n$ scaling of pure-state simulation. Consequently, the noisy simulation was executed using Qiskit Aer (*GenericBackendV2*) via the PennyLane-Qiskit interface, while the noiseless runs used the *lightning.qubit* state-vector simulator.

The noisy HQ-PINN training converged successfully with a total runtime of approximately 668 seconds, using 1024 shots per circuit evaluation. In quantum computing, a shot refers to one independent execution (or sampling) of a quantum circuit. Because quantum measurements are probabilistic, multiple shots are required to estimate expectation values accurately. Increasing the number of shots improves statistical precision but proportionally increases computation time. The noisy simulation produced physically consistent inversion results (Fig. 9) but with higher misfits than the noiseless case- 0.5352 km $s^{-1}$ g $cm^{-3}$ for acoustic impedance and 0.000983 for seismic data- reflecting the expected effects of decoherence and gate errors on circuit expressivity and optimization stability. While noisy simulators provide a realistic approximation of NISQ-era behaviour, real hardware may display somewhat different timing characteristics. Physical devices can execute circuits in parallel on dedicated quantum control electronics, potentially reducing per-iteration runtime relative to density-matrix simulators, which emulate noise classically. However, hardware queue delays and limited access can offset these advantages. Hence, the inclusion of noisy simulations provides valuable insight into the robustness of the HQ-PINN framework under realistic operating conditions.

## Discussion

This paper explores a novel application of QML in seismic inversion, laying the groundwork for integrating quantum computing into geophysical workflows. Specifically, we demonstrate how quantum computing frameworks can be effectively combined with classical machine learning to implement seismic inversion, estimating P- and S-impedances from post-stack and pre-stack seismic data using Hybrid Quantum Physics-Informed Neural Networks (HQ-PINNs).

In all case studies, quantum circuits were implemented using the PennyLane library, which facilitates hybrid quantum-classical computations on various quantum hardware and simulators. It is important to note that similar results can also be achieved using other quantum programming platforms, such as Qiskit and Cirq, although terminologies like feature maps, ansatz, and their pre-defined templates may differ across platforms. For this study, we utilized the amplitude embedding feature map to exploit the power of quantum superposition in encoding high-dimensional seismic data. This method enables an exact encoding of classical amplitudes into quantum states but often at the cost of deeper circuits and higher gate counts. For very high-dimensional problems, variational encoding may be used to reduce this computational overhead, though with some loss of accuracy. Alternatively, angle embedding can offer greater hardware efficiency, particularly for near-term quantum devices when the input seismic data is not very high dimensional.

For the synthetic case studies, the HQ-PINN model was trained for 500 epochs. As the model trained rapidly, training was not truncated before reaching 500 epochs. The learning curves indicate that convergence was achieved, and similar results were obtained after just 250 epochs, suggesting that practical applications could further reduce training time by truncating earlier. For the quantum node of the HQ-PINN model, we primarily employed the basic entangler layers ansatz. We also tested the strongly entangling layers ansatz, which incorporates Controlled-X, Controlled-Y, and Controlled-Z gates. This configuration achieved similar accuracy but required slightly longer training times- on the order of a few additional seconds. Therefore, for clarity and computational efficiency, we present the results obtained using the basic entangler layer ansatz in this study.

The HQ-PINN model achieved reliable impedance predictions even when trained on noisy simulators, underscoring its robustness and signifying a critical step toward realizing HQ-PINN deployment on actual quantum hardware. The landscape of quantum hardware is rapidly evolving, and various platforms hold unique potential for advancing QML applications in geophysics. Superconducting qubits (e.g., IBM Quantum, Rigetti, Google Quantum AI), photonic quantum systems (e.g., Xanadu, PsiQuantum), neutral atoms (e.g., QuEra, Pasqal), and trapped ions (e.g., IonQ, Honeywell Quantum Solutions) are all promising technologies. Each hardware platform comes with its distinct advantages: superconducting qubits excel in scalability and gate fidelity, photonic systems offer room-temperature operation and speed, neutral atoms provide large-scale highly configurable qubit arrays, and trapped ions deliver long coherence times and precise control. In future work, we plan to test HQ-PINNs across these platforms to leverage their technological advancements and evaluate their suitability for geophysical applications. Furthermore, we aim to extend our proposed framework to Full Waveform Inversion (FWI), addressing the limitations of classical FWI workflows and unlocking new possibilities for high-resolution subsurface imaging.

## Conclusions

Quantum computing holds transformative potential for addressing computationally intensive problems across various scientific domains, and geophysics is no exception. By leveraging the principles of quantum mechanics, quantum computers have the potential to significantly enhance efficiency in solving high-dimensional processing and imaging problems, which are prevalent in geophysical workflows. We introduce hybrid quantum physics-informed neural networks (HQ-PINNs) and demonstrate their efficacy across synthetic and field case studies for pre- and post-stack seismic inversion. In the HQ-PINN model, the quantum layer serves as the input layer and consists of an amplitude embedding feature map, basic entangler layers (ansatz) with $\boldsymbol{R_X}, \boldsymbol{R_Y}$ or $\boldsymbol{R_Z}$ rotation gates, and measurements in the computational basis. Notably, the choice of $\boldsymbol{R_X}, \boldsymbol{R_Y}$ or $\boldsymbol{R_Z}$ gates yielded accurate and similar results. Interestingly, even the no-ansatz configuration (i.e., no trainable parameters in the quantum layer) produced accurate predictions, underscoring the effectiveness of the embedding layer. This capability to encode seismic data in the quantum Hilbert space demonstrates the QNode's potential as a robust dimensionality reduction tool for solving high-dimensional imaging problems. The performance of the HQ-PINN model is influenced by the quantum device and the differentiation method used to train the quantum layer. For smaller circuits, the *lightning.qubit* simulator was faster than the *lightning.gpu* device, primarily due to reduced overhead and efficient CPU-based optimizations. However, as the circuit size increased, the GPU's ability to handle larger state spaces in parallel become evident, as seen with the *lightning.gpu* device offering significant speedups for the Sleipner inversion case for joinly inverting sections from two surveys with a 17-qubit configuration. Among differentiation methods, SPSA and adjoint emerged as the most efficient. SPSA achieves its speed by approximating the gradient through simultaneous perturbation of all parameters, requiring only two function evaluations per iteration regardless of the parameter count. The adjoint method is also efficient, as it computes the gradient using a single forward pass and a backward pass, leveraging the circuit structure for scalability. Finite difference is slower because it estimates the gradient by perturbing each parameter individually, resulting in multiple function evaluations for each parameter, which becomes computationally expensive in high-dimensional parameter spaces. The parameter-shift rule requires evaluating the quantum circuit at multiple shifted parameter values per parameter, leading to significant computational overhead, but it offers analytical gradients and is compatible with quantum hardware. Thus, parameter-shift is typically slower than adjoint and SPSA but may outperform finite difference depending on implementation and hardware. In the Sleipner field case study, we performed both individual and simultaneous inversions of two baseline surveys to estimate acoustic impedances. Remarkably, the simultaneous inversion required only one additional logical qubit to invert two 2D seismic sections, leveraging the power of quantum superposition. This result highlights the potential of HQ-PINNs as a highly efficient framework for addressing data-intensive geophysical problems. Even the no-ansatz configuration produced accurate results, further emphasizing the effectiveness of quantum embedding for projecting classical data into high-dimensional quantum Hilbert spaces. When noisy simulators were used to mimic real quantum hardware, the HQ-PINN model continued to produce accurate impedance estimates, albeit with longer training times due to mixed-state simulations and the need for multiple-shot evaluations. This experiment demonstrates the model's robustness under realistic noise conditions and lays the groundwork for its eventual implementation on actual quantum hardware. The HQ-PINN framework is scalable, flexible, and adaptable to various inverse problems, quantum programming frameworks, and classical automatic differentiation libraries. By demonstrating the efficacy and efficiency of HQ-PINNs across both synthetic and field case studies, this work not only underscores the potential of hybrid QML for geophysical applications but also establishes a foundation for future research in quantum-enhanced geosciences.

## Acknowledgments

We acknowledge the Stanford Energy Science and Engineering Department and the sponsors of the Stanford Center for Earth Resources Forecasting (SCERF) for their support in conducting this research. We would also like to thank Equinor for sharing the Sleipner benchmark model.

## Appendix A: Quantum Computing and Quantum Machine Learning

In this section, we provide a brief overview of the theoretical foundations of quantum computing and quantum machine learning. For a more comprehensive discussion, we refer readers to Nielsen and Chuang (2010) and Schuld and Petruccione (2021). Qubits are the fundamental units of information in quantum computing, analogous to bits in classical computing. However, unlike classical bits, which can only exist in one of two definite states (0 or 1), qubits can exist in a superposition of these states. This superposition is mathematically represented in Dirac notation (or bra-ket notation) as:

$$|\boldsymbol{\psi}\rangle = \boldsymbol{\alpha}|\mathbf{0}\rangle + \boldsymbol{\beta}|\mathbf{1}\rangle, \qquad (\mathbf{A1})$$

where $|\boldsymbol{\psi}\rangle$ is a vector in a Hilbert space, and $|\mathbf{0}\rangle$ and $|\mathbf{1}\rangle$ are the basis vectors, represented in matrix form as:

$$|\mathbf{0}\rangle = \begin{bmatrix} \mathbf{1} \\ \mathbf{0} \end{bmatrix}, |\mathbf{1}\rangle = \begin{bmatrix} \mathbf{0} \\ \mathbf{1} \end{bmatrix}. \qquad (\boldsymbol{A2})$$

The coefficients $\boldsymbol{\alpha}$ and $\boldsymbol{\beta}$ are complex numbers that determine the probability of measuring the qubit in a particular state, with the normalization condition: $|\boldsymbol{\alpha}|^{\mathbf{2}} + |\boldsymbol{\beta}|^{\mathbf{2}} = \mathbf{1}$ ensuring that the total probability sums to 1. According to the Born rule, when a qubit is measured, its state collapses to one of the basis states, $|\mathbf{0}\rangle$ or $|1\rangle$, with probabilities $|\boldsymbol{\alpha}|^{\mathbf{2}}$ and $|\boldsymbol{\beta}|^{\mathbf{2}}$ respectively. This collapse is an irreversible process due to quantum measurement. To perform desired computations, the state of a qubit must be manipulated and this is achieved using quantum gates.

Quantum gates are operations (unitary transformations) that manipulate qubits by altering their probability amplitudes and relative phases. These gates are mathematically represented by unitary matrices, meaning they preserve the norm of the qubit's state vector, ensuring that quantum operations are reversible. Mathematically, a unitary matrix $\boldsymbol{U}$ satisfies the condition $\boldsymbol{U}^{\dagger}\boldsymbol{U} = \boldsymbol{U}\boldsymbol{U}^{\dagger} = \boldsymbol{I}$ where $\boldsymbol{U}^{\dagger}$ is the conjugate transpose (Hermitian adjoint) of $\boldsymbol{U}$, and $\boldsymbol{I}$ is the identity matrix. A sequence of quantum gates applied to a set of qubits constitutes a quantum circuit, which serves as the computational backbone of quantum algorithms.

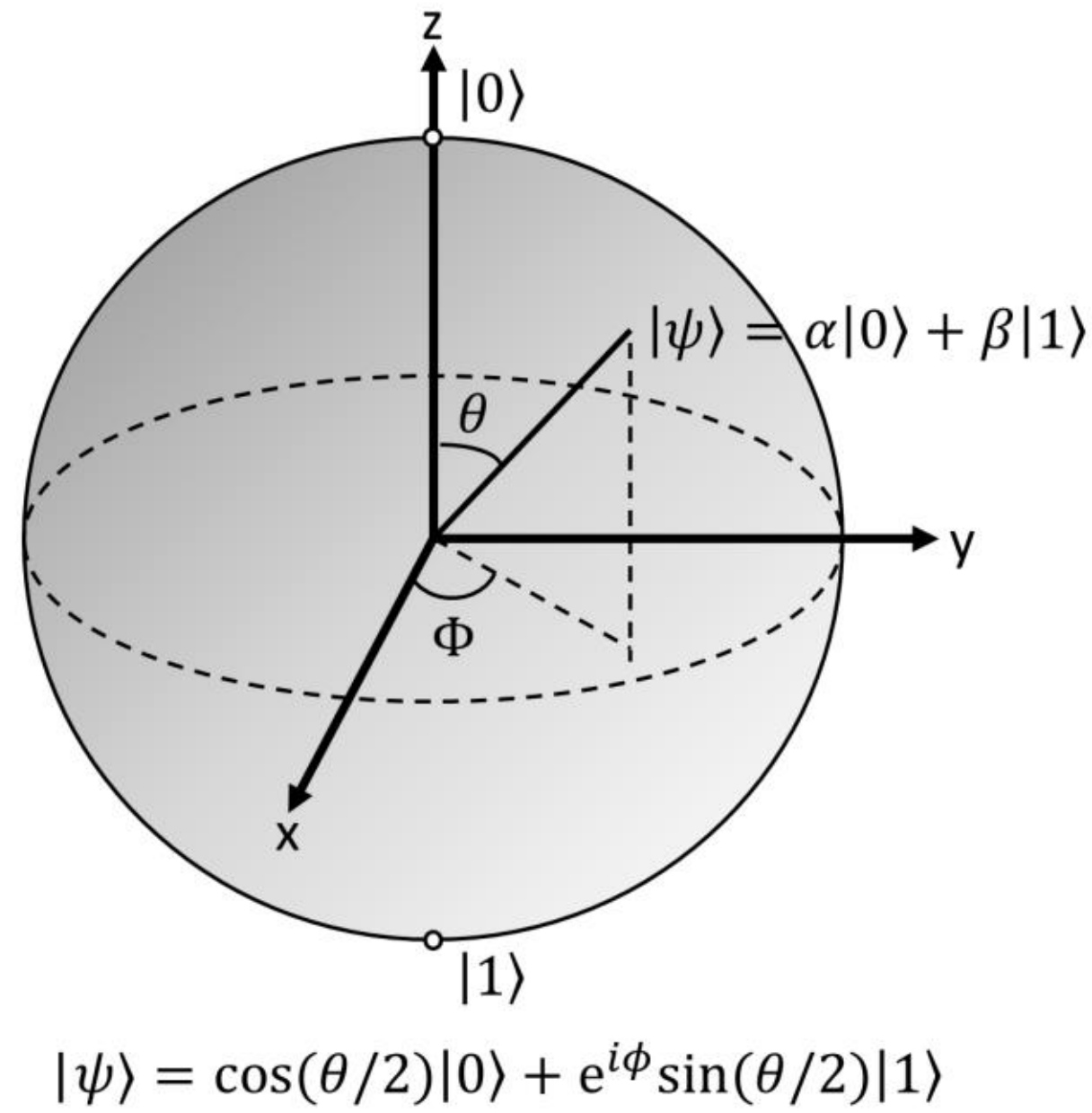

**Figure A1**. The Bloch sphere illustrating the quantum state of a single qubit $|\psi\rangle$ in the Hilbert space. Quantum gate operations manipulate $|\psi\rangle$ by performing unitary transformations, which correspond to rotations of the state vector around the Bloch sphere.

Single-qubit gates can be visualized as rotations on the Bloch sphere (Fig. A1), a geometrical representation where the state of a qubit is depicted as a point on a unit sphere in three-dimensional space. The qubit's state can be parametrized using spherical coordinates $\boldsymbol{\theta}$ and $\boldsymbol{\phi}$, expressed as:

$$|\boldsymbol{\psi}\rangle = \mathbf{cos}(\boldsymbol{\theta}/\mathbf{2})|\mathbf{0}\rangle + \boldsymbol{e^{i\phi}}\mathbf{sin}(\boldsymbol{\theta}/\mathbf{2})|\mathbf{1}\rangle. \qquad (\mathbf{A3})$$

Here, $\boldsymbol{\theta}$ (polar angle) determines the relative weight (probability distribution) of $|\mathbf{0}\rangle$ and $|\mathbf{1}\rangle$, while $\boldsymbol{\phi}$ (azimuthal angle) introduces a relative phase between the basis states. This representation emphasizes the qubit's complex probabilistic nature and its evolution through quantum gates. Quantum gates correspond to rotations of the state vector $|\boldsymbol{\psi}\rangle$ on the Bloch sphere, allowing precise control over qubit states. A general single-qubit rotation gate $\boldsymbol{R}(\boldsymbol{\theta}, \boldsymbol{\phi}, \boldsymbol{\lambda})$ is defined as:

$$\boldsymbol{R}(\boldsymbol{\theta}, \boldsymbol{\phi}, \boldsymbol{\lambda}) = \begin{bmatrix} \boldsymbol{cos}\dfrac{\boldsymbol{\theta}}{\mathbf{2}} & -\boldsymbol{e^{i\lambda}}\,\boldsymbol{sin}\dfrac{\boldsymbol{\theta}}{\mathbf{2}} \\ \boldsymbol{e^{i\phi}}\,\boldsymbol{sin}\dfrac{\boldsymbol{\theta}}{\mathbf{2}} & \boldsymbol{e^{i(\phi+\lambda)}}\,\boldsymbol{cos}\dfrac{\boldsymbol{\theta}}{\mathbf{2}} \end{bmatrix}. \qquad (\mathbf{A4})$$

This general rotation gate enables flexible transformations of qubit states, forming the basis for more complex quantum operations. Some commonly used quantum gates include the rotation gates ($\boldsymbol{R_X(\theta)}, \boldsymbol{R_Y(\theta)}, \boldsymbol{R_Z(\theta)}$}, the Pauli gates ($\boldsymbol{\widehat{X}}, \boldsymbol{\widehat{Y}}$ and $\boldsymbol{\widehat{Z}}$), and the Hadamard Gate ($\boldsymbol{H}$). These gates have been briefly explained in the Appendix B.

When dealing with multiple qubits, their states and operations are represented using tensor products. For a system with $\boldsymbol{n}$ qubits, the combined quantum state is described as a vector in a $\boldsymbol{2^n}$-dimensional Hilbert space. For example, the joint state of two qubits, $|\boldsymbol{\psi_1}\rangle$ and $|\boldsymbol{\psi_2}\rangle$, is represented as $|\boldsymbol{\psi_1}\rangle \otimes |\boldsymbol{\psi_2}\rangle = |\boldsymbol{\psi_1\psi_2}\rangle$, where $\otimes$ denotes the tensor product. If $|\boldsymbol{\psi_1}\rangle = \boldsymbol{\alpha}|\mathbf{0}\rangle + \boldsymbol{\beta}|\mathbf{1}\rangle$ and $|\boldsymbol{\psi_2}\rangle = \boldsymbol{\gamma}|\mathbf{0}\rangle + \boldsymbol{\delta}|\mathbf{1}\rangle$, the combined state is given by: $|\boldsymbol{\psi_1\psi_2}\rangle = (\boldsymbol{\alpha}|\mathbf{0}\rangle + \boldsymbol{\beta}|\mathbf{1}\rangle) \otimes (\boldsymbol{\gamma}|\mathbf{0}\rangle + \boldsymbol{\delta}|\mathbf{1}\rangle) = \boldsymbol{\alpha\gamma}|\mathbf{00}\rangle + \boldsymbol{\alpha\delta}|\mathbf{01}\rangle + \boldsymbol{\beta\gamma}|\mathbf{10}\rangle + \boldsymbol{\beta\delta}|\mathbf{11}\rangle$. The probabilities of the states $|\mathbf{00}\rangle$, $|\mathbf{01}\rangle$, $|\mathbf{10}\rangle$, $|\mathbf{11}\rangle$ are given by $|\boldsymbol{\alpha\gamma}|^2$, $|\boldsymbol{\alpha\delta}|^2$, $|\boldsymbol{\beta\gamma}|^2$, $|\boldsymbol{\beta\delta}|^2$ respectively, and they sum to one.

Similarly, quantum gates applied to multi-qubit systems are represented as tensor products of individual gates. An illustration is demonstrated in Appendix C. Like single-qubit gates, multi-qubit gates also exist in quantum computing. The Controlled-NOT (CNOT) gate for example, is a two-qubit gate where the state of one qubit (the control qubit) determines whether the other qubit (the target qubit) is flipped. The CNOT gate is a key component in creating entangled states and forms the basis of many quantum algorithms. The CNOT gate flips the target qubit if the control qubit is $|\mathbf{1}\rangle$. If the control qubit is $|\mathbf{0}\rangle$, the target qubit remains unchanged. The CNOT gate is given as:

$$\boldsymbol{CNOT} = \begin{bmatrix} \mathbf{1} & \mathbf{0} & \mathbf{0} & \mathbf{0} \\ \mathbf{0} & \mathbf{1} & \mathbf{0} & \mathbf{0} \\ \mathbf{0} & \mathbf{0} & \mathbf{0} & \mathbf{1} \\ \mathbf{0} & \mathbf{0} & \mathbf{1} & \mathbf{0} \end{bmatrix}. \qquad (\mathbf{A5})$$

$$\boldsymbol{CNOT}|\mathbf{10}\rangle = |\mathbf{11}\rangle\,. \qquad (\boldsymbol{A6})$$

In addition to the CNOT gate, quantum circuits often use controlled rotation gates, which apply a rotation operation to a target qubit only when the control qubit is in the $|\mathbf{1}\rangle$ state. These include controlled-$\boldsymbol{R_X}$, controlled-$\boldsymbol{R_Y}$ and controlled-$\boldsymbol{R_Z}$ gates. In general, any unitary matrix $\boldsymbol{U}$ can be converted into a controlled-$\boldsymbol{U}$ gate. Additionally, it is also possible for controlled operations to be conditioned on the control qubit being in the $|\mathbf{0}\rangle$ state.

Quantum gates can also be parameterized to introduce tunable parameters, providing the flexibility needed to optimize quantum circuits for specific tasks. These parameterized quantum gates form the foundation of Parameterized Quantum Circuits (PQCs), where the parameters are adjusted dynamically to perform targeted quantum transformations. A PQC operates by applying a series of parameterized unitary transformations $\boldsymbol{U(\theta)}$ to an input quantum state $|\boldsymbol{\psi_{in}}\rangle$, yielding an output state: $|\boldsymbol{\psi_{out}}\rangle = \boldsymbol{U(\theta)}|\boldsymbol{\psi_{in}}\rangle$, where $\boldsymbol{U(\theta)}$ depends on a vector of parameters $\boldsymbol{\theta}$, which are optimized during computation. The structure of a PQC is guided by the choice of ansatz, a circuit template designed to explore the Hilbert space efficiently. The ansatz must be expressive yet computationally feasible to enable effective optimization. Commonly used ansatzes include the hardware-efficient ansatz (Kandala et al., 2017; Leone et al., 2024; Xiao et al., 2024), which is tailored to the physical capabilities of quantum devices, and problem-inspired ansatzes (Peruzzo et al., 2014; Choquette et al., 2021; Wang et al., 2024a), which encode domain-specific knowledge.

Another critical component of quantum machine learning is the feature map, which encodes classical data into quantum states by embedding it within the quantum Hilbert space (Schuld and Killoran, 2019; Havlíček et al., 2019). This mapping process applies a sequence of quantum gates to prepare an initial quantum state that represents the input data. Two widely used encoding methods are angle embedding and amplitude embedding:

1) Angle Embedding: This approach encodes each classical data point $\boldsymbol{x_i}$ as the rotation angle of a quantum gate. For a classical data vector $\boldsymbol{x} = [\boldsymbol{x_1}, \boldsymbol{x_2}, \dots, \boldsymbol{x_n}]$, the corresponding quantum state is:

$$|\boldsymbol{\psi}(\boldsymbol{x})\rangle = \bigotimes_{i=1}^{n} \boldsymbol{R}_{\{X,Y,Z\}}(\boldsymbol{x_i})|\boldsymbol{0}\rangle, \qquad \textbf{(A7)}$$

where $\boldsymbol{R}_{\{X,Y,Z\}}(\boldsymbol{x_i})$ represents a single-qubit rotation gate (see Appendix B). Since quantum rotation gates are periodic, $\boldsymbol{x_i}$ must be preprocessed (e.g., scaled to $[\boldsymbol{0}, \boldsymbol{\pi}]$) to preserve uniqueness.

2) Amplitude Embedding: In this method, classical data is encoded directly into the amplitudes of a quantum state. For a classical data vector: $\boldsymbol{x} = [\boldsymbol{x_1}, \boldsymbol{x_2}, \dots, \boldsymbol{x_N}]$, the corresponding quantum state is:

$$|\boldsymbol{\psi}(\boldsymbol{x})\rangle = \frac{\boldsymbol{1}}{\|\boldsymbol{x}\|} \sum_{i=1}^{N} \boldsymbol{x_i}|\boldsymbol{i}\rangle, \qquad \textbf{(A8)}$$

where $|\boldsymbol{i}\rangle$ denotes the computational basis states, $\|\boldsymbol{x}\|$ ensures normalization ($\sum|\boldsymbol{x_i}|^2 = \boldsymbol{1}$), and $\boldsymbol{N} = \boldsymbol{2^n}$ (if $\boldsymbol{N}$ is not a power of 2, zero-padding is applied), requiring $\mathbf{log_2(N)}$ qubits.

To encode floating-point data, such as a seismic trace, into the amplitudes of a quantum state $|\boldsymbol{\psi_{seis}}\rangle$, algorithms like the Möttönen et al. (2005) state preparation method can be used to determine the architecture of the quantum embedding circuit. This algorithm constructs a quantum circuit that transforms the initial state $|\boldsymbol{00} \dots \boldsymbol{0}\rangle$ into an arbitrary quantum state $|\boldsymbol{\psi_{seis}}\rangle$ by applying a sequence of controlled rotations. Specifically, controlled-$\boldsymbol{R_Z}$ gates adjust the phases, while controlled-$\boldsymbol{R_Y}$ gates adjust the amplitudes. The required rotation angles are

computed based on the phases and amplitudes of $|\boldsymbol{\psi}_{seis}\rangle$. A numerical demonstration of this algorithm is given in Example 4.1 of Chapter 4 in Schuld and Petruccione (2021).

Angle embedding is generally advantageous for smaller datasets because it relies on single-qubit rotations, making it hardware-efficient and less prone to noise. However, it does not fully exploit entanglement, which may limit its expressiveness for complex data distributions. Amplitude embedding, while more resource-intensive, offers higher data capacity by encoding data directly into the amplitudes of a quantum state. This approach is better suited for large datasets but requires deeper circuits and more controlled operations, making it challenging for near-term quantum hardware. The choice between these methods depends on dataset size, algorithmic requirements, and the available quantum hardware.

After embedding the classical dataset into the quantum Hilbert space, parameterized gates within the ansatz act on the quantum state. This process transforms the embedded state into a form that, upon measurement, provides valuable information for the learning task. The output of the quantum circuit is obtained by measuring the expectation value of a Hermitian operator $\widehat{\boldsymbol{O}}$, typically one of the Pauli operators like $\widehat{\boldsymbol{Z}}$ or $\widehat{\mathbf{X}}$. The expectation value is given by:

$$\langle\widehat{\boldsymbol{O}}\rangle = \langle\boldsymbol{\psi}(\boldsymbol{\theta},\boldsymbol{x})|\widehat{\boldsymbol{O}}|\boldsymbol{\psi}(\boldsymbol{\theta},\boldsymbol{x})\rangle\,. \qquad \textbf{(A9)}$$

Here $\langle\widehat{\boldsymbol{O}}\rangle$ represents the expected measurement outcome, and $|\boldsymbol{\psi}(\boldsymbol{\theta},\boldsymbol{x})\rangle$ is the output quantum state for a given input $\boldsymbol{x}$ and parameters $\boldsymbol{\theta}$.

Quantum neural networks (QNNs) leverage this mechanism to perform machine learning tasks by optimizing the parameters $\boldsymbol{\theta}$ of the ansatz. This is achieved by defining a loss function $\boldsymbol{\mathcal{L}}(\boldsymbol{\theta})$ which measures the difference between the predicted and target outcomes. The expectation values of the qubits serve as the output of the quantum layer, which in case of HQNNs can be passed to subsequent classical layers (with parameters $\boldsymbol{w}$) for further processing. The loss function ($\boldsymbol{\mathcal{L}}(\boldsymbol{\theta})$ for QNNs and $\boldsymbol{\mathcal{L}}(\boldsymbol{\theta},\boldsymbol{w})$ for HQNNs) is then calculated and the objective is to minimize this loss, ensuring the model's parameters $\boldsymbol{\theta}$ and $\boldsymbol{w}$ are optimized for accurate predictions.

Optimization of the parameters $\boldsymbol{\theta}$ in quantum layers is typically carried out by gradient-based techniques to adjust the quantum circuit's parameters to minimize the loss function. Several methods can be used to compute this gradient, including the parameter-shift rule, adjoint method, finite-difference method, and simultaneous perturbation stochastic approximation (SPSA). A brief overview of these techniques is provided in Appendix D.

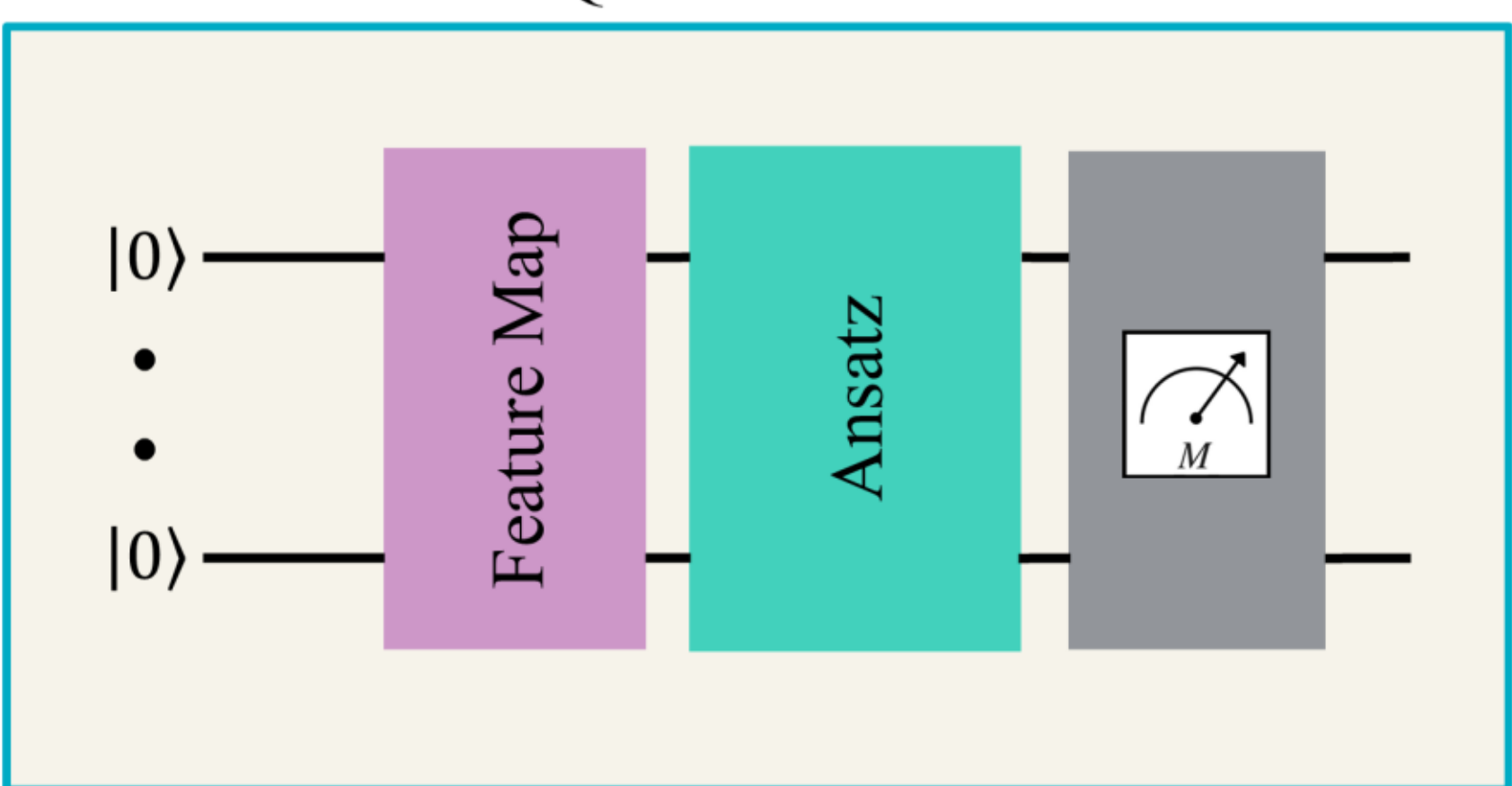

**Figure A2**. A Quantum Node (QNode) comprising feature map, ansatz, and measurement operators. The feature map encodes classical data into quantum states, the ansatz employs parameterized quantum gates to process the encoded data, and the measurement operators extract meaningful classical information from the quantum states.

Automatic differentiation further facilitates seamless propagation of gradients of the loss function with respect to both $\boldsymbol{\theta}$ and $\boldsymbol{w}$ through the hybrid quantum models, enabling efficient end-to-end optimization. The development of such frameworks necessitates robust software tools. Several open-source platforms, including Qiskit by IBM (Javadi-Abhari et al., 2024), Cirq (Google, 2020) by Google, and PennyLane (Bergholm et al., 2018) by Xanadu, provide comprehensive interfaces for designing and simulating quantum circuits, as well as integrating them with classical machine learning frameworks. Among these, PennyLane offers a flexible framework for QML and seamlessly integrates with widely-used classical machine learning libraries such as PyTorch (Paszke et al., 2019) and TensorFlow (Abadi et al., 2016). A key feature of PennyLane is the Quantum Node, or QNode (Fig. A2), which encapsulates a quantum circuit and enables automatic differentiation by integrating the quantum circuit with the computational graphs of classical neural networks, facilitating their gradient-based optimization within classical machine learning workflows. PennyLane also provides access to high-performance quantum simulator devices like *lightning.qubit* and *lightning.gpu*. The *lightning.qubit* device is a statevector simulator written in C++ and optimized for CPU execution, supporting the differentiation methods for efficient gradient computations in large quantum circuits. The *lightning.gpu* device is also a statevector simulator that leverages NVIDIA's cuQuantum SDK for GPU-accelerated simulations, offering significant speedups for deep and complex circuits. In this study, we utilize PennyLane in conjunction with PyTorch to design and implement the HQ-PINN framework (Fig. 1). We leverage its predefined templates for feature maps, ansatz circuits, simulator devices, and differentiation techniques to perform seismic inversion.

## Appendix B: Commonly used Single-Qubit Gates

1) The Rotation gates $\{\boldsymbol{R_X(\theta)}, \boldsymbol{R_Y(\theta)}, \boldsymbol{R_Z(\theta)}\}$: The rotation gates $\boldsymbol{R_X(\theta)}, \boldsymbol{R_Y(\theta)}$, and $\boldsymbol{R_Z(\theta)}$, perform rotations of a qubit state around the $\boldsymbol{X}, \boldsymbol{Y}$ and $\boldsymbol{Z}$ axes of the Bloch sphere by an angle $\boldsymbol{\theta}$, respectively. They are fundamental in quantum computing for precise qubit state manipulation and are mathematically represented as:

$$\boldsymbol{R_X(\theta) = \begin{bmatrix} cos\left(\frac{\theta}{2}\right) & -i\,sin\left(\frac{\theta}{2}\right) \\ -i\,sin\left(\frac{\theta}{2}\right) & cos\left(\frac{\theta}{2}\right) \end{bmatrix}}, \qquad (\mathbf{B1a})$$

$$\boldsymbol{R_Y(\theta) = \begin{bmatrix} cos\left(\frac{\theta}{2}\right) & -sin\left(\frac{\theta}{2}\right) \\ sin\left(\frac{\theta}{2}\right) & cos\left(\frac{\theta}{2}\right) \end{bmatrix}}, \qquad (\mathbf{B1b})$$

$$\boldsymbol{R_Z(\theta) = \begin{bmatrix} e^{-i\frac{\theta}{2}} & 0 \\ 0 & e^{i\frac{\theta}{2}} \end{bmatrix}}. \qquad (\mathbf{B1c})$$

When $\boldsymbol{\theta = \pi}$, special cases arise for all three gates:

$$\boldsymbol{R_X(\pi) = \begin{bmatrix} 0 & -i \\ -i & 0 \end{bmatrix} = -i\begin{bmatrix} 0 & 1 \\ 1 & 0 \end{bmatrix} = -i\widehat{X}}, \qquad (\mathbf{B2a})$$

$$\boldsymbol{R_Y(\pi) = \begin{bmatrix} 0 & -1 \\ 1 & 0 \end{bmatrix} = -i\begin{bmatrix} 0 & -i \\ i & 0 \end{bmatrix} = -i\widehat{Y}}, \qquad (\mathbf{B2b})$$

$$\boldsymbol{R_Z(\pi) = \begin{bmatrix} -i & 0 \\ 0 & i \end{bmatrix} = -i\begin{bmatrix} 1 & 0 \\ 0 & -1 \end{bmatrix} = -i\widehat{Z}}. \quad (\mathbf{B2c})$$

$\boldsymbol{\widehat{X}}, \boldsymbol{\widehat{Y}}$ and $\boldsymbol{\widehat{Z}}$ are called the Pauli-X, Y and Z gates respectively. The global phase $\boldsymbol{-i}$ in each Pauli gate arises because the rotation gates are derived from unitary transformations. Global phases $\boldsymbol{e^{i\phi}}$ uniformly affect all components of the quantum state and do not influence the measurement probabilities or observable outcomes. Thus, the equivalence between the Pauli gates and their respective rotation gates (up to a global phase) is valid for all practical purposes.

The Pauli-X gate plays a fundamental role in quantum computing, acting as a quantum equivalent of the classical NOT gate. It flips $|\mathbf{0}\rangle$ to $|\mathbf{1}\rangle$ and vice versa. The Pauli-Y gate combines the action of the Pauli-X gate (bit flip) with a phase flip, making it useful in certain quantum algorithms that require combined transformations, such as quantum error correction. The Pauli-Z gate is crucial for modifying the relative phase of quantum states between $|\mathbf{0}\rangle$ and $|\mathbf{1}\rangle$, which is key to quantum interference and entanglement operations.

2) Hadamard Gate ($\boldsymbol{H}$): The Hadamard gate is a single-qubit gate that creates superposition by transforming the computational basis states $|\mathbf{0}\rangle$ and $|\mathbf{1}\rangle$ into an equal superposition of these states. It is one of the most fundamental quantum gates in quantum computing, used to initialize qubits in a superposition state.

$$\boldsymbol{H} = \frac{\mathbf{1}}{\sqrt{\mathbf{2}}}\begin{bmatrix} \mathbf{1} & \mathbf{1} \\ \mathbf{1} & -\mathbf{1} \end{bmatrix}. \qquad (\mathbf{B3})$$

$$\boldsymbol{H}|\mathbf{0}\rangle = \frac{\mathbf{1}}{\sqrt{\mathbf{2}}}(|\mathbf{0}\rangle + |\mathbf{1}\rangle). \qquad (\mathbf{B4})$$

## Appendix C: Illustration of Single-Qubit Gate Operations in Multi-Qubit Quantum Circuits

Consider the application of a bit-flip operator or the Pauli-$\boldsymbol{X}$ gate to the first qubit of a two-qubit system initially in the state $|\mathbf{00}\rangle$. The operation is represented as $\boldsymbol{X} \otimes \boldsymbol{I}$, where $\boldsymbol{I}$ is the identity gate that leaves the second qubit unchanged.

Let the initial state of the system be:

$$|\mathbf{00}\rangle = |\mathbf{0}\rangle \otimes |\mathbf{0}\rangle = \begin{bmatrix} \mathbf{1} \\ \mathbf{0} \end{bmatrix} \otimes \begin{bmatrix} \mathbf{1} \\ \mathbf{0} \end{bmatrix} = \begin{bmatrix} \mathbf{1} \\ \mathbf{0} \\ \mathbf{0} \\ \mathbf{0} \end{bmatrix}. \qquad (\mathbf{C1})$$

The operator can then be represented as:

$$\boldsymbol{X} \otimes \boldsymbol{I} = \begin{bmatrix} \mathbf{0} & \mathbf{1} \\ \mathbf{1} & \mathbf{0} \end{bmatrix} \otimes \begin{bmatrix} \mathbf{1} & \mathbf{0} \\ \mathbf{0} & \mathbf{1} \end{bmatrix} = \begin{bmatrix} \mathbf{0} & \mathbf{0} & \mathbf{1} & \mathbf{0} \\ \mathbf{0} & \mathbf{0} & \mathbf{0} & \mathbf{1} \\ \mathbf{1} & \mathbf{0} & \mathbf{0} & \mathbf{0} \\ \mathbf{0} & \mathbf{1} & \mathbf{0} & \mathbf{0} \end{bmatrix}. \qquad (\mathbf{C2})$$

Applying $\boldsymbol{X} \otimes \boldsymbol{I}$ to $|\mathbf{00}\rangle$, we get:

$$\begin{bmatrix} \mathbf{0} & \mathbf{0} & \mathbf{1} & \mathbf{0} \\ \mathbf{0} & \mathbf{0} & \mathbf{0} & \mathbf{1} \\ \mathbf{1} & \mathbf{0} & \mathbf{0} & \mathbf{0} \\ \mathbf{0} & \mathbf{1} & \mathbf{0} & \mathbf{0} \end{bmatrix} \begin{bmatrix} \mathbf{1} \\ \mathbf{0} \\ \mathbf{0} \\ \mathbf{0} \end{bmatrix} = \begin{bmatrix} \mathbf{0} \\ \mathbf{0} \\ \mathbf{1} \\ \mathbf{0} \end{bmatrix} = \begin{bmatrix} \mathbf{0} \\ \mathbf{1} \end{bmatrix} \otimes \begin{bmatrix} \mathbf{1} \\ \mathbf{0} \end{bmatrix} = |\mathbf{1}\rangle \otimes |\mathbf{0}\rangle, \qquad (\mathbf{C3})$$

which corresponds to the state $|10\rangle$, showing that the $X$ gate has flipped the first qubit while leaving the second qubit unchanged.

## Appendix D: Overview of Differentiation Methods for Quantum Layers

1) Parameter-Shift Rule: This provides an analytical technique for calculating gradients of expectation values with respect to circuit parameters. For a parameter $\boldsymbol{\theta}$, the gradient is determined as:

$$\frac{\partial\langle\widehat{\boldsymbol{O}}\rangle}{\partial\boldsymbol{\theta}} = \frac{\mathbf{1}}{\mathbf{2}}\left[\langle\widehat{\boldsymbol{O}}\left(\boldsymbol{\theta}+\frac{\boldsymbol{\pi}}{\mathbf{2}}\right)\rangle - \langle\widehat{\boldsymbol{O}}\left(\boldsymbol{\theta}-\frac{\boldsymbol{\pi}}{\mathbf{2}}\right)\rangle\right], \qquad (\mathbf{D1})$$

where $\langle\widehat{\boldsymbol{O}}(\boldsymbol{\theta})\rangle$ is the expectation value of the observable $\widehat{\boldsymbol{O}}$ for a given parameter $\boldsymbol{\theta}$. This method is applicable to many quantum gates and is relatively straightforward to implement. However, it necessitates multiple evaluations of the quantum circuit for each parameter, which can become computationally expensive, particularly for complex circuits with numerous parameters.

2) Adjoint Method: This is an efficient technique for computing gradients, particularly useful for simulating large quantum circuits. It calculates gradients by reversing the quantum operations in the circuit, utilizing the adjoint (conjugate transpose) of the unitary gates. The gradient of an expectation value $\langle\widehat{\boldsymbol{O}}\rangle$ is computed by applying the adjoint of the unitary transformation $\boldsymbol{U}(\boldsymbol{\theta})$, allowing for all gradients to be computed in a single backward pass instead of requiring separate evaluations for each parameter. This approach is computationally efficient, as it eliminates the need for multiple circuit runs for each parameter. However, it requires hardware or software capable of executing circuits in reverse, which is currently not feasible on most quantum hardware.

3) Finite Difference Method: For cases where analytical differentiation is impractical, this method serves as a numerical alternative for estimating gradients. This approach perturbs the parameter $\boldsymbol{\theta}$ by a small increment $\boldsymbol{\delta}$ and approximates the gradient as:

$$\frac{\partial\langle\widehat{\boldsymbol{O}}\rangle}{\partial\boldsymbol{\theta}} \approx \frac{\langle\widehat{\boldsymbol{O}}(\boldsymbol{\theta}+\boldsymbol{\delta})\rangle - \langle\widehat{\boldsymbol{O}}(\boldsymbol{\theta}-\boldsymbol{\delta})\rangle}{\mathbf{2}\boldsymbol{\delta}}. \qquad (\mathbf{D2})$$

Although straightforward, this method can suffer from inefficiencies due to numerical approximations and the need for multiple circuit evaluations. Additionally, its performance may degrade in the presence of noise.

4) Simultaneous Perturbation Stochastic Approximation (SPSA): This is a gradient estimation method particularly well suited for high-dimensional quantum optimization problems. In this approach, all parameters are perturbed simultaneously in a stochastic manner, and the gradient is estimated as:

$$\frac{\partial\langle\hat{O}\rangle}{\partial\theta_i} \approx \frac{\langle\hat{O}(\vec{\theta} + \epsilon\vec{\Delta})\rangle - \langle\hat{O}(\vec{\theta} - \epsilon\vec{\Delta})\rangle}{2\epsilon\Delta_i}, \qquad (D3)$$

where $\vec{\Delta}$ is a vector of random perturbations applied to all parameters, and $\epsilon$ is a small scalar. SPSA is computationally efficient for optimizing large parameter spaces but may require more iterations to achieve convergence due to its stochastic nature.